\documentclass[preprint,11pt]{elsarticle}

\usepackage{color,soul}
\usepackage{gensymb}
\usepackage{amsmath,amssymb,bm}
\usepackage{mathtools}
\usepackage{nomencl}
\usepackage{soul}
\usepackage{graphicx,subfig,float}
\usepackage{geometry}
\usepackage{pdflscape,rotating}
\usepackage[normalem]{ulem}

\newcommand{\ON}{$\mathcal{O}(N)$ }
\newcommand{\ONLOGN}{$\mathcal{O}(N \log{} N)$ }
\DeclarePairedDelimiter\abs{\lvert}{\rvert}%
\DeclarePairedDelimiter\norm{\lVert}{\rVert}%
\DeclarePairedDelimiter\p{\lparen}{\rparen}
\makeatletter
\let\oldabs\abs
\def\abs{\@ifstar{\oldabs}{\oldabs*}}
\let\oldnorm\norm
\def\norm{\@ifstar{\oldnorm}{\oldnorm*}}
\makeatother

\journal{tp be submitted to Computational Physics}

\makenomenclature

\begin{document}

\begin{frontmatter}

\title{Error-Controlled Hybrid Adaptive Fast Solver for Regularized Vortex Methods}

\author[AUB]{Samer Salloum}
\author[AUB]{Issam Lakkis\corref{cor}}
\cortext[cor]{Corresponding author Tel: 961350000, ext 3636, Email: il01@aub.edu.lb}
\address[AUB]{Department of Mechanical Engineering, American University of Beirut, Beirut, Lebanon}

\begin{abstract}
In this paper, an error-controlled hybrid adaptive fast solver that combine both \ON and \ONLOGN scheme is proposed. For a given accuracy, the adaptive solver is used in the context of regularized vortex methods to optimize the speed of the velocity and vortex stretching calculation. This is accomplished by introducing three critical numbers in order to limit the depth of the tree division and to balance the near-field and far-field calculations for any hardware architecture. The adaptive solver is analyzed in term of speed and accuracy.
\end{abstract}


\end{frontmatter}

\newpage

\section{Introduction}

Vortex methods are mesh-free methods used to solve the Navier-Stokes equation by tracking the motion of vorticity-carrying fluid elements. In these methods, the calculation of the velocity and vortex stretching vectors at each element location is done through direct summation over all vortex elements. Thus for $\cal N$ particles simulation, the direct evaluation of the all pairs interaction requires $O{(\cal N }^2 )$ computational operations. For large $ N$, the cost become prohibitive even with the use of fast computers \cite{6127848}. This high cost led scientist to develop approximate solutions for the $\cal N$ body problems.\\
Historically, scientists have developed two variants of hierarchical $\cal N $-body methods, Treecodes and FMMs \cite{DBLP:journals/corr/abs-1209-3516}. The Treecode algorithm \cite{Barnes:1986uq} reduces the complexity to \ONLOGN by clustering source particles into progressively larger groups and using multipole expansions to approximate their influence on each target particle. The FMM \cite{Greengard:1987kx} can achieve \ON by clustering not only the source particles, but also the nearby target particles using local expansions to approximate their influence at the targets locations\cite{6127848}. These two methods  have followed very different paths of evolution and differs from each other in several way \cite{article,dehnen2002hierarchical,6127848,yokota2011parameter}:
\begin{enumerate}
\item For the far field, Treecodes performs cell-particle interactions leading to \ONLOGN complexity, while FMM performs cell-cell interactions leading to \ON complexity
\item Treecodes have conventionally used hard coded cartesian expansions with a fixed order of expansion $p = 3$  while FMMs have used spherical harmonic expansions of an arbitrary order
\item Treecodes use the ratio between the size of cells and the distance between them to construct the interaction list. This is known as the multipole acceptance criterion (MAC), and it is used to determine if a cell should be evaluated or subdivided further. In contrast, the FMM uses parent, child, and neighbor relationships to construct the interaction list \cite{6127848}
\item In Treecodes, the accuracy is controlled by the MAC parameter $\theta$, while in FMMs, accuracy is controlled by  the order of expansion $p$
\item In Treecodes, the expansions are centered on the cells' centers of mass rather than geometrical centers in FMMs \cite{dehnen2002hierarchical}
\item Treecodes are adaptive in nature and permits interaction of cells at different level. FMMs employ a (usually) non-adaptive structure of hierarchical grids and considers only interactions between nodes on the same grid level according to their geometrical neighborhood \cite{article}
\end{enumerate}


There have been some efforts from both the Treecode and FMM  community to take the best of these two methods and produce an optimal algorithm \cite{yokota2011parameter}. In 1995, Michael Warren and John Salmon \cite{WARREN1995266} developed a new technique to calculate cell-cell interactions using Taylor series in Cartesian coordinates. 
Yet they didn't implement cell-cell interaction in the Treecode framework. In 1999, Cheng et al. \cite{CHENG1999468} extended the work of Greengard \cite{Greengard:1987kx} and introduced a more adaptive cell-cell interaction stencil that considers the interaction of cells at different levels in the tree. They also introduced a mechanism to select between cell-particle and cell-cell interactions in FMMs, which is always faster than a pure Treecode or pure FMM. In 2000, Walter Dehnen \cite{article} extended the Treecode method introduced by Barnes and hut \cite{Barnes:1986uq} for fast evaluation of gravitational forces by including mutual cell-cell interaction developed in \cite{WARREN1995266}. He also introduced a symmetric multipole-acceptance criterion (MAC) and use it to determine whether a given cell-cell interaction can be executed or the cell must be split. In 2002, Walter Dehnen \cite{dehnen2002hierarchical} presented an improvement of his original code by introducing other techniques such as generic tree traversals, mutual cell-cell interactions, and a mass-dependent error-controlled multipole acceptance criteria. 
Walter code is divided into two phases, namely  the interaction phase and the evaluation phase. In the interaction phase, for each cell, the Taylor series coefficients of all its interactions (cell that interact via Multipole criteria) are evaluated and accumulated. 
In the evaluation phase, for each body, the Taylor series of all relevant cells (those that contain the body) are accumulated by first translating to a common expansion center and then adding coefficients. Walter compared his code directly with the 3D adaptive FMM code by Cheng et al. \cite{CHENG1999468}. On average his code is faster by more than a factor of 10 and twice as accurate . However, his code cannot compete with the Cheng et al. FMM in the regime E $(10^{-6})$. Walter concluded that traditional FMM is less useful in the low-accuracy regime, such as is needed in stellar dynamics, in agreement with earlier findings \cite{capuzzo1998comparison}. He argued that the reason is that his code increases accuracy by decreasing the MAC number ($p$ is always constant ) while FMM increases accuracy by increasing $p$. He finished by saying that a code for which $p$ and  $\theta$ can be adapted simultaneously would be superior to both codes. In 2012, Yokota and Barba \cite{6127848} developed a hybrid Treecode-FMM algorithm that has a similar structure to Dehnen's method and has control over both the order of expansion and MAC. Their code is based on spherical harmonic expansions, but have the capability to switch to Cartesian expansions if the required accuracy is lower than a certain threshold. In addition they incorporate the capability to auto-tune the kernels on heterogeneous architectures. The key for auto-tuning is to time all the kernels. The algorithm uses this information to select the optimum kernel during the dual tree traversal, choosing between cell-cell, cell-particle, and particle-particle interactions. Yokota and Barba tested their code by fixing the maximum number of elements per leaf to 50 and 200 respectively. For a well chosen critical number ($N=200$), the results indicate that the hybrid method always favors cell-cell interactions and doesn't provide a visible advantage over the pure FMM. However for ($N=50$) the FMM suffers from load imbalance between the near-field and far-field evaluations, while the hybrid method doesn't because it can choose to perform particle-particle interactions even if the cell is not at the leaf level. They concluded that the hybrid method produced a better result by automatically fine tuning the balance between the particle-particle and cell-cell interactions throughout the adaptive tree. It optimizes the balance between particle-particle and cell-cell interactions and achieves optimum performance for all $ N$. There's no need to tweak parameters (such as particles per cell) to achieve optimal performance. The hybrid method seems to remove the wavy behavior of the $\cal N$-dependence in pure FMMs \cite{yokota2011fast}. In 2013, Yokota and Barba \cite{yokota2011parameter} investigated the effect of changing both the multiple acceptance criteria $\theta$ and the order of expansion $p$ to achieve maximum performance for a given accuracy on both CPUs and GPUs. They have shown that it is never efficient to increase $\theta$ to a value that is larger than that of standard FMM codes ($\theta = 1/2$), for both CPU and GPU implementations. Therefore, when designing a hybrid Treecode-FMM algorithm it is not necessary to have a variable MAC which would have an optimum value of $1/2$. In this case, the hybrid scheme will produce similar interaction list as the one described by Cheng et al. (1999), but the implementation for the hybrid scheme is much simpler and the algorithm itself is more general and flexible. \\

\noindent The hybrid scheme developed by Yokota and Barba was able to tune the balance between the particle-particle and cell-cell interaction when FMM suffers from load imbalance. By using pre-measured kernel, Yokota and Barba also manage to solve the hardware dependence problematic when heterogeneous architectures come into the picture. However, the fact that the maximum number of elements per cell is introduced by the user will pose some problems. If the maximum number of elements per leaf was much higher than the optimal number, Yokota and Barba scheme won't be able to balance cell-cell and particle-particle interactions. In fact, although the hybrid scheme will always favors cell-cell interactions for non-neighboring cells, the particle to particle interaction will take most of the simulation time since we have a large number of elements within neighboring cells. The interaction between neighboring cells cannot be evaluated using approximate methods since this will introduce an unbounded errors. Moreover, if the maximum number of elements per leaf was much smaller than the optimal number, and since Yokota did not introduce any mechanism to limit the depth of the tree structure, the hybrid scheme will manage to balance cell-cell and particle-particle interaction, but part of the simulation time will be spent to evaluate multipole and local expansions at the center of cell of high level, although these multipole and local expansions are not needed to evaluate the approximate field.\\
On the other hand, Cheng did not give any particular attention to different hardware architectures and he didn't present any effective mechanism to limit the Oct-tree depth. In his paper \cite{CHENG1999468}, Cheng introduced four different adaptive list associated with each cell and uses particle-particle interaction if the number of elements in  a given source cell is less than $p^2$. Moreover, the maximum number of elements per cell is also introduced by the user. This will yield the same problems described in the hybrid scheme case.\\

In treecodes and FMMs, the maximum number of particles per cell $N_{crit}$ is set (by the user) to balance the loads of near-field and far-field evaluations. Setting this number to be too small will result in a deep tree structure with a disproportionately large amount of far-field and too few near-field evaluations. Conversely, if $N_{crit}$ is set to be too large, the resulting tree structure will be too shallow and a large amount of time will be spent in the near-field evaluation \cite{6127848}. In this paper, we introduce two different critical numbers, $n_T$ and $n_F$.  These critical numbers are used in the adaptive Cheng FMM framework to limit the depth of the tree structure in the upward and downward pass. A variant of these two numbers are also used to determine the type of interactions that need to be evaluated in order to tune the balance between  cell-cell, cell-particle and particle-particle interactions. These two critical numbers are evaluated as a function of the ratio of time pre-measured kernel and differs from one architecture to another, thus we manage to solve the hardware dependence problematic in Cheng algorithm. The user does not need to worry about the maximum number of elements per leaf because it is chosen automatically by the code to ensure FMM balance. These critical numbers vary with the order of expansion $p$, the choice of the cutoff function, the vector fields that need to be evaluated at each time step and the hardware architecture. For a given level of accuracy, which is associated with an order of expansion $p$, the adaptive solver evaluate $n_T$ and $n_F$ and uses these two critical numbers to minimize the amount of computational work needed to evaluate the truncated singular expressions of the velocity and vortex stretching vectors. However, the use of a regularization core function will introduce an additional error since these same expansions are used to approximate the far-field regularized velocity and vortex stretching vectors. This additional error is related to choice of the cutoff function and mainly depends on the size of the smallest box obtained upon tree division. Hence, and in order to maintain the level of accuracy in regularized vortex methods, we introduce a critical level $l_c$, which is equal to the maximumum level that can be reached upon tree division for a given level of accuracy.

\section{Vortex methods and the Vorticity Transport equation}\label{sec:vorticityTransportEquation}
For three dimensional incompressibe flow, $\nabla .\vec{u} = 0$, the evolution of the vorticity of a fluid particle is described by the Helmholtz's vorticity equation:
\begin{equation}
\frac{d\vec{\omega}}{dt}=( \vec{\omega} .\nabla)\vec{u}+\frac{\nabla \rho \times \nabla P}{\rho}+\nabla^2 \vec{\omega}
\label{eq:Helmholtz's equation}
\end{equation}

For an incompressible flow in unbounded domain, the velocity vector field $\vec{u}$ in equation \ref{eq:Helmholtz's equation}, is computed as the curl of a vector potential field $\vec{\psi}$
\begin{equation}
\vec{u}=\nabla\times\vec{\psi}
\end{equation}
The vorticity vector field $\vec{\omega}$ in equation \ref{eq:Helmholtz's equation}, is computed as the curl of the velocity vector field
\begin{equation}
\vec{\omega}=\nabla\times\vec{u}=\nabla\times(\nabla\times\vec{\psi})
\end{equation}
so that $\vec{\psi}$ solves the Poisson equation
\begin{equation}
\nabla^2 \vec{\psi}=- \vec{\omega}
\label{eq:poisson}
\end{equation}

Noting that the Green's function for the Poisson equation (\ref{eq:poisson}) in
an unbounded domain is $G(\vec{x})=-\frac{1}{4\pi \parallel\vec{x}\parallel}$ we obtain
\begin{equation}
\vec{\psi}(\vec{x},t)=\frac{1}{4\pi}\int{\frac{\vec{\omega}(\vec{y},t)}{\norm{\vec{x}-\vec{y}}} d^3y}
\end{equation}
\\

Assuming a singular representation of the vorticity field,
\begin{equation}
  \vec\omega _{\delta}(\vec{x},t) = \sum_i \vec{\alpha}_i \delta (\vec{x}-\vec{x}_i),
  \label{eq:singularVorticityRepresentation}
\end{equation}
the vetor potential field and the velocity vector field can be approximated as
\begin{equation}
\vec{\psi}_{\delta}(\vec{x},t) = \frac{1}{4\pi}\sum_{i}\frac{\vec{\alpha}_i(t)}{\norm{\vec{x}(t)-\vec{x}_i (t)}}
\label{eq:streamlineDirectCalculation}
\end{equation}
\begin{equation}
\vec{u}_{\delta}(\vec{x},t) =  \nabla\times\vec{\psi}_{\delta} =-\frac{1}{4\pi}\sum_i\frac{\left(\vec{x}(t)-\vec{x}_i(t)\right)}{\norm{\vec{x}(t)-\vec{x}_i (t) }^3}\times\vec{\alpha}_i(t)
\label{eq:velocityDirectCalculation}
\end{equation}

Assuming a regularized representation of the vorticity field,
\begin{equation}
\vec\omega_\sigma (\vec{x},t)=\zeta_\sigma (\vec{x})*\vec{\omega} (\vec{x},t)=\sum_{i}\vec{\alpha}_i (t)\zeta_\sigma (\vec{x}-\vec{x}_i )
\label{eq:regularizedVorticityRepresentation}
\end{equation}
the vector potential field and the velocity vector field can be approximated as
\begin{equation}
\vec{\psi}_\sigma (\vec{x},t)=\sum_{i}G_\sigma (\vec{x}-\vec{x}_i )\vec{\alpha}_i (t)
\label{eq:streamlineBlobCalculation}
\end{equation}
\begin{equation}
\vec{u}_{\sigma}(\vec{x},t)= \nabla\times\vec{\psi}_{\sigma}=-\sum_i q_{\sigma}\left(\vec{x}(t)-\vec{x}_i (t)\right) \frac{(\vec{x}(t)-\vec{x}_i(t))}{\norm{\vec{x}(t)-\vec{x}_i (t) }^3}\times\vec{\alpha}_i(t)
\label{eq:velocityBlobCalculation}
\end{equation}
$\zeta_\sigma$ is a radially symmetric regularization function and and $\sigma$ is its associated smoothing radius, $G$ solves the Poisson equation  $\nabla^2 G =-\zeta (r) $, $G_\sigma (r) = \frac{1}{\sigma} G \p{\frac{r}{\sigma}}$, $q(r)=\int_{0}^r \zeta(\rho)\rho^2 d\rho$, and $q_{\sigma}(r)=q \p{\frac{r}{\sigma}}$ \\
\\

To accurately simulate three dimensional flows, a very large number particles is required to resolve the various length scales. Direct evaluation of the velocity and vortex stretching vectors requires $ O(N^2 )$ computational operations of all the pairwise interaction in a system of $N$ particles. The prohibitive cost of the direct evaluation renders the fast multipole method (FMM) an essential tool to execute particle-based simulations in a reasonable amount of time. 

\section{The Fast Multipole in three dimensions - Adaptive Scheme}\label{sec:fastN}

The computational cost of the $\cal N$ body problem is tremendously reduced by arranging nearby source particles into a recursively smaller boxes and using multipole expansion coefficients $\vec{M}_n^m$ in order to approximate the far field components of the velocity and vortex stretching vectors induced by all sources within the source box. Further reduction is achieved by clustering nearby target elements, in addition to source elements, and using local expansion coefficients $\vec{L}_n^m$ to approximate the aforementioned vectors at any location within a target box induced by all distant sources in the computational domain. The expansion is based on spherical harmonics $Y_n^m$.\\

\begin{figure}[!htb]
\begin{center}
\includegraphics[width=5in]{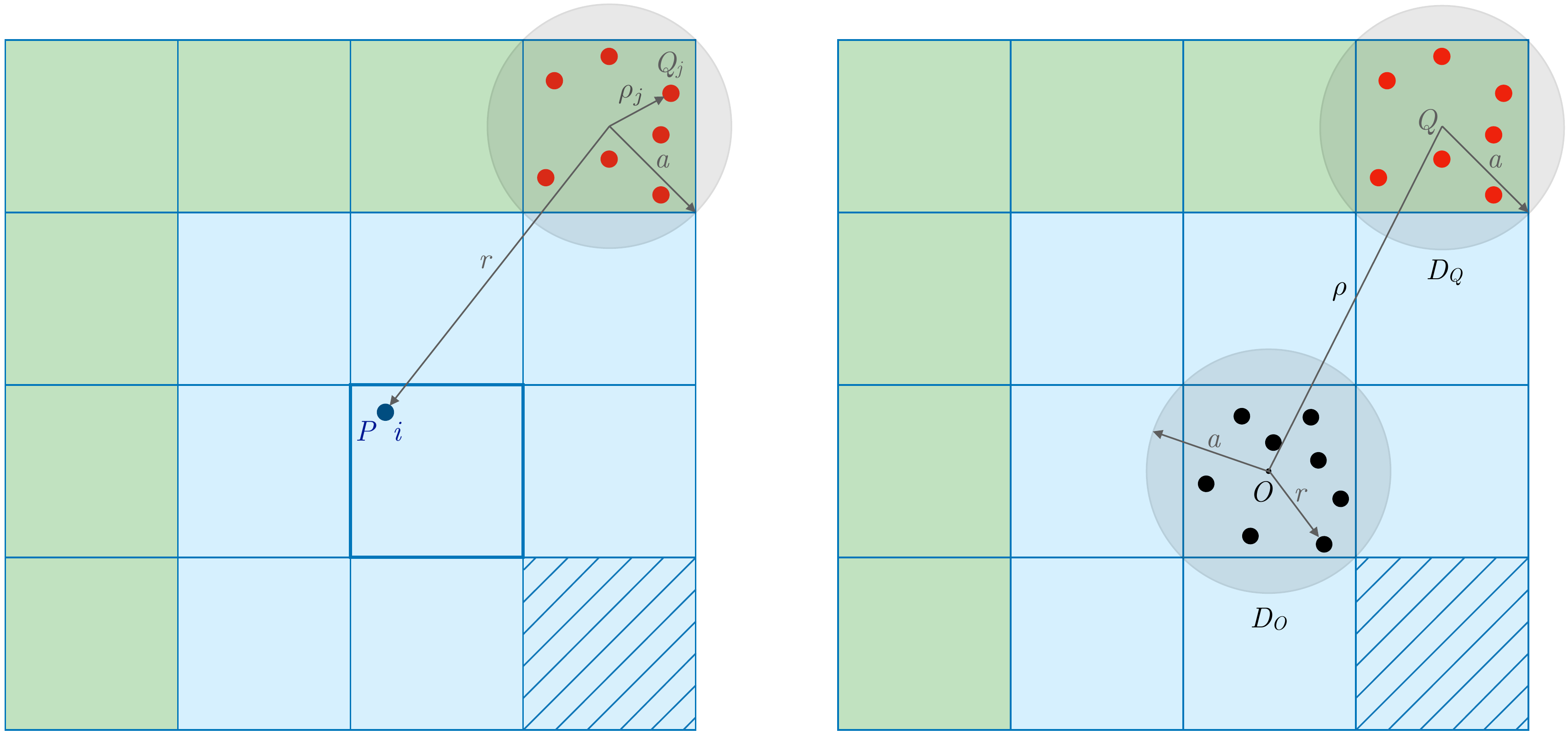}
\caption{Schematic for the multipole and local expansions.}
\label{schematicFMMNandLOGN}
\end{center}
\end{figure}

\noindent Assuming $s$ vortices with strength $a$ $(\vec{\alpha}_j, j=1...s)$ located at the points $\vec{Q}_j=(\rho_j,\theta_j,\varphi_j)$ inside the sphere $D_Q$, with $\abs{\rho_j} < a$ (see Figure \ref{schematicFMMNandLOGN}, left), then at any $\vec{P}=(r,\theta,\varphi)$ with $r>a$, the vector potential is approximated by the following multipole expansion
\begin{equation}
\vec{\psi}_{\delta}(\vec{P},t) \simeq \frac{1}{4\pi}\sum_{n=0}^p \sum_{m=-n}^n \frac{\vec{M}_n^m}{r^{n+1}} Y_n^m (\theta,\varphi)
\label{eq:multipoleExpansionApproximation}
\end{equation}
with
\begin{equation}
\vec{M}_n^m=\sum_{j=1}^s \vec{\alpha}_j\rho_j^n Y_n^{-m} (\theta_j,\varphi_j)
\label{eq:calculatingMultipoleExpansionCoefficients}
\end{equation}
where the coordinates of the target point $\vec{P}=(r,\theta,\varphi)$ are taken with respect to an orthonormal system centered at $Q$.

Furthermore, suppose that $s$ vortices having strengths $ (\vec{\alpha}_j, j=1...s)$ are located inside the sphere $D_Q$ of radius $a$ with center at $Q=(\rho,\alpha,\beta)$ (see Figure \ref{schematicFMMNandLOGN}, right), and that $\rho = (c+1)a $ with $c >1$, then for any point $ P(r,\theta,\phi) $ inside $D_0$ of radius $a$ centered at the origin, the vector potential due to vortices $ (\vec{\alpha}_j, j=1...s)$ inside $D_Q$ is approximated by the following local expansion
\begin{equation}
\vec{\psi}_{\delta}(P,t) \simeq \frac{1}{4\pi}\sum_{n=0}^p \sum_{m=-n}^n \vec{L}_n^m Y_n^m (\theta,\varphi)r^n,
\label{eq:localExpansion}
\end{equation}
where
\begin{equation}
\vec{L}_n^m =\sum_{j=0}^\infty \sum_{k=-j}^j \frac{\vec{M}_j^k.i^{\arrowvert m-k \arrowvert-\arrowvert m \arrowvert-\arrowvert k \arrowvert}.A_j^k .A_n^m.\rho^j .Y_{n+j}^{k-m}(\alpha,\beta)}{(-1)^j A_{n+j}^{k-m}.\rho^{n+j+1}},
\label{eq:Conversion}
\end{equation}
with $A_n^m$ defined by
\begin{equation}
A_n^m=\frac{(-1)^n}{\sqrt{(n-m)!(n+m)!}}
\end{equation}
\\

Simple expressions for the velocity and vortex stretching vectors in term of multipole and local expansion coefficients are derived in \cite{}. These expressions are used in the context of an hybrid adaptive fast multipole scheme to approximate the far-field vectors.\\ 

\begin{figure}[!htb]
\begin{center}
\includegraphics[width=5in]{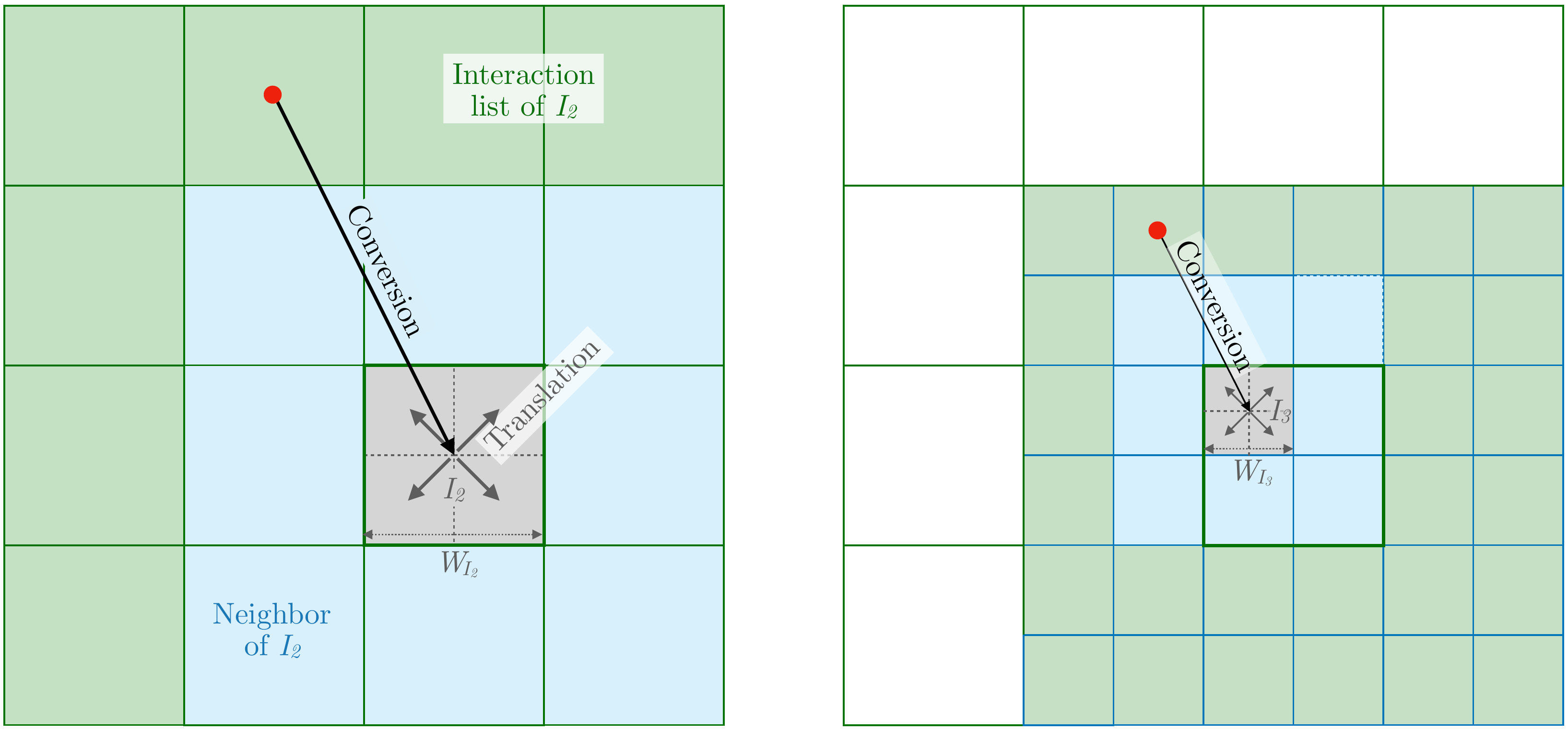}
\caption{Schematic for the \ON FM scheme.}
\label{schematicFMMN}
\end{center}
\end{figure}

In order to describe and analyse the adaptive solver, we will use the following notation:
\begin{itemize}
\item ${\cal N}_{n_D}(I_l)$: For a given target box $I_l$ at level $l$, a source box $J_l$ at the same level belong to the neighborhood of $I_l$ , ${\cal N}_{n_D}(I_l)$ , if and only if the distance along $x$, $y$, and $z$ between the center of the two boxes is less than or equal to $n_D W_{J_l}$ where $n_D$ is a positive integer and $W_{J_l}$ is the width of box $J_l$
\item ${\cal T}_{n_D}(I_l)$: For a given target box $I_l$ at level $l$, the Interaction List ${\cal T}_{n_D}(I_l)$ , is the set of all boxes which are children of Neighbors of $I_l$'s parent and are not neighbors of $I_l$ ( green boxes in Figure \ref{schematicFMMN}). For $n_D =1$, we will have similar interaction list as the one described by Cheng et al. (1999) and by the hybrid scheme desribed by Yokota for an optimum MAC value of $\frac{1}{2}$
\item ${\cal H}_{n_D}(I_l)$: For a given target box $I_l$ at level $l$, the Inherited List ${\cal H}_{n_D}(I_l)$, is the set of all boxes that belong to the interaction list of $I_l$ ancestor, and for which the adaptive solver choose not perform cell to cell interaction. As her name suggest, this list is transmitted from a parent box to its children.
\item ${\cal C}_{n_D}(I_l)$: For a given target box $I_l$ at level $l$, ${\cal C}_{n_D}(I_l)$, is the set of all boxes that belong to the interaction list of $I_l$ ancestor, and for which the adaptive solver choose to evaluate their influence through cell to cell interaction
\item ${\cal M}_{n_D}(I_l)$: For a given target box $I_l$ at level $l$, ${\cal M}_{n_D}(I_l)$, is the set of all source boxes for which the adaptive solver chooses to evaluate their influence through cell to particle interaction
\end{itemize}
\newpage
\begin{figure}[!htb]
\begin{center}
\includegraphics[width=2.5in]{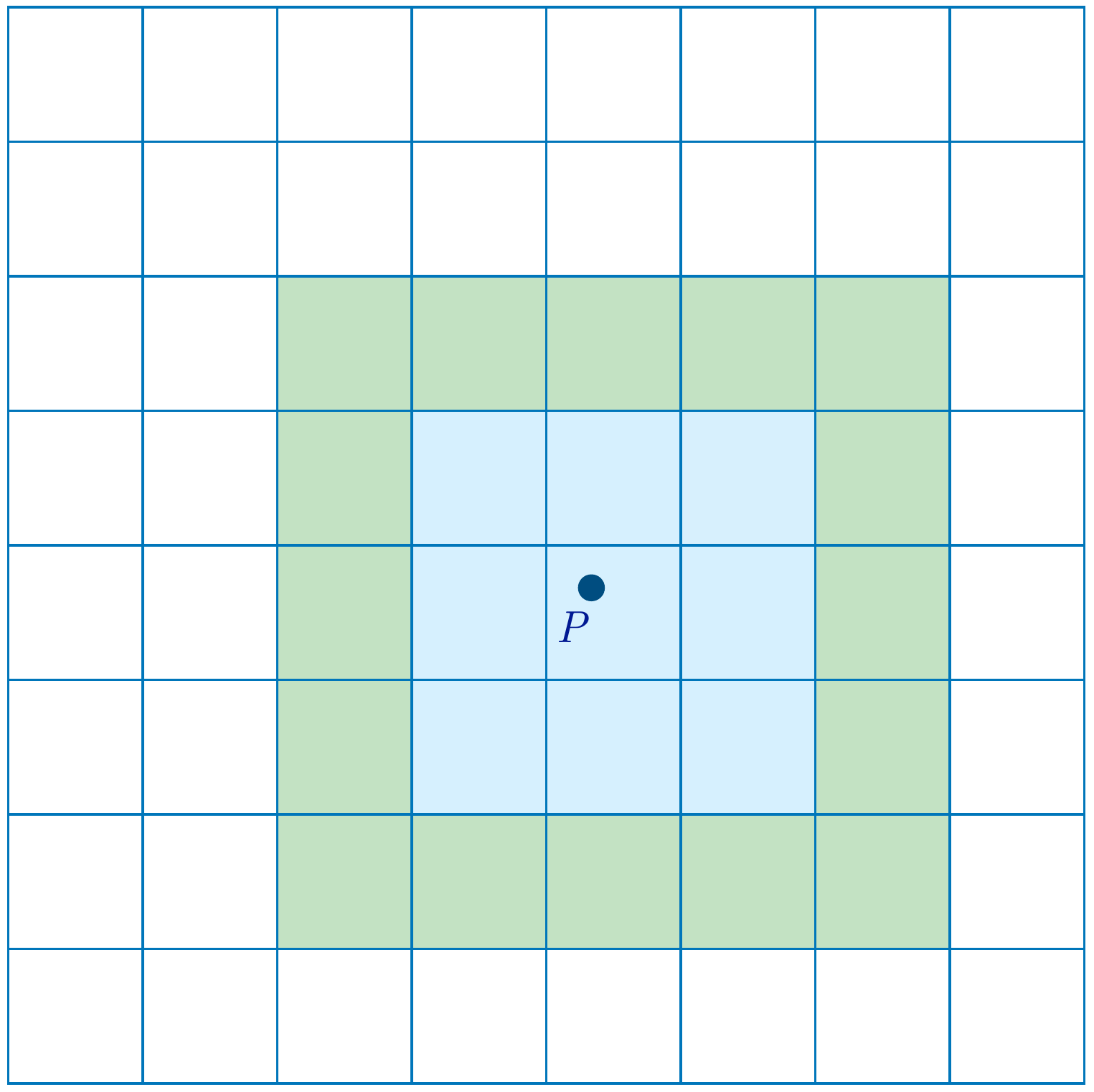}
\caption{Neighborhood ${\cal N}(j), {\cal N}(J)$ for $n_D=1$ (blue) and $n_D=2$ (green)}
\label{schematicNeighborhood}
\end{center}
\end{figure}

\begin{figure}[!htb]
\begin{center}
\includegraphics[width=5in]{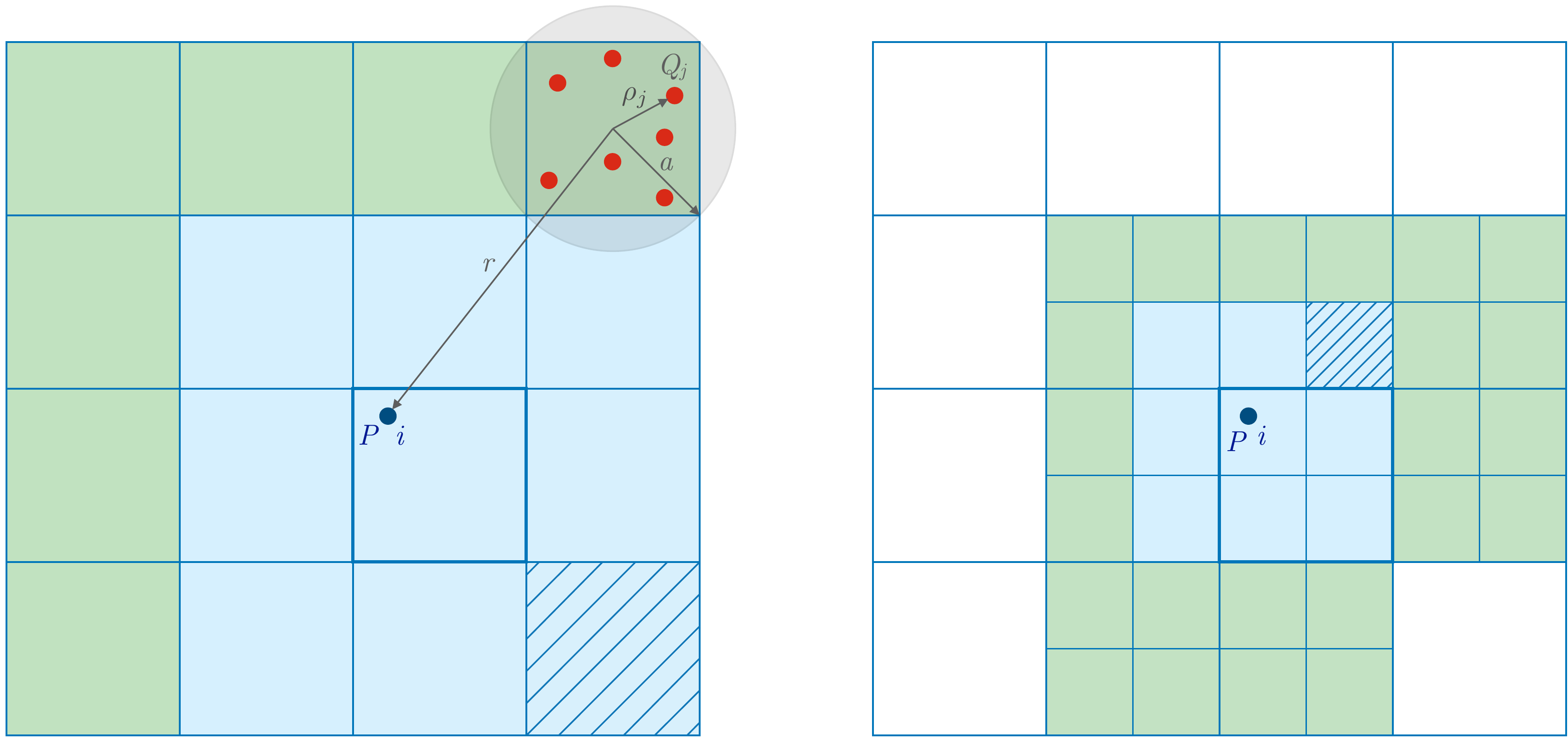}
\caption{Schematic for the \ONLOGN FM scheme.}
\label{schematicFMMNLOGN}
\end{center}
\end{figure}

The adaptive scheme consists of the following steps:

\begin{enumerate}
\item Construct the smallest cubic box that contains all the vortex elements in the computational domain. This box is called the root box.
\item Construct a hierarchy of boxes by recursively dividing the root box uniformly into smaller boxes.
Refinement level $0$ is equivalent to the root box. Refinement level $l+1$ is obtained by subdivision of each box at level $l$ into eight equally sized children. The division process stops whenever the box level $l$ is equal to the critical level $l_c$ or the number of elements in the box is less $n_C = \min \{n_T,n_F \}$, where $n_T$ and $n_F$ are defined in section \ref{sec:costAnalysis}. A box that has no children is called a leaf box.
\item Initial expansion: At the finest level, all sources are expanded at their box centers to obtain the far-field multipole expansion coefficients $\vec{M}_n^m $ using equation (\ref{eq:calculatingMultipoleExpansionCoefficients}).
\item Upward pass: The multipole expansion coefficients for each box are translated via multipole-to-multipole translations from the source box centers to their parent source box center. All these translations are performed in a hierarchical order from bottom to top via the octree.
\item Downward pass: Form local expansion coefficients at the center of each box at every mech level $ l \geq 2 $. These local expansion coefficients are calculated by the following recursive manner: local expansion coefficients for all boxes at level $1$ are first set to zero. For any target box $I_l$ at level $ l \geq 2 $ that contains $N_T$ targets, initial local expansion coefficients are obtained by translating local coefficients from the center of $I_l$ parent to $I_l$ center. For any Source box $J_l$ that belongs to the interaction list of $I_l$ ( green boxes in Figure \ref{schematicFMMN}) and contains $N_S$ sources, if $N_T$ is greater than $\frac{n_F^2}{8n_T}$ and $N_T * N_S$ is greater than $\frac{n_F^2}{64}$, the multipole coefficients at the center of $J_l$ are converted into local expansion coefficients at the center of box $I_l$ using equation (\ref{eq:Conversion})and then added to the initial local coefficients. Otherwise, $J_l$ is set in the Inherited list of $I_l$. The downward pass stops whenever the number of elements in a given box is less than a critical number $n_{F}$ defined in section \ref{sec:costAnalysis}
\item Calculate the velocity and vortex stretching vectors: For every target element $i$ in the computational domain, identify leaf box $I_{l^*}$ that contains this element. 
\begin{itemize}
\item Calculate the far-field velocity and vortex stretching vectors at the target position using $I_{l^*}$ local expansion coefficients.
\item Calculate the velocity and vortex stretching vectors at the target position due to all sources contained within $I_{l^*}$ neighbours and Inherited lists using the following recursive manner, which is identical to the recursive method used in the \ONLOGN scheme:
\begin{itemize}
\item  For each box $J \in {\cal N}_{n_D}(I_l)$ or ${\cal H}_{n_D}(I_l)$, if the number of elements in $J$ is less than $\frac{n_T}{8}$ the velocity and vortex stretching vectors induced at $P$ by all source elements contained in $J$ are calculated by direct summation, else if the target element $i$ is not contained in $J$ nor in its neighbors (green boxes in Figure \ref{schematicFMMNLOGN}), the velocity and vortex stretching vectors induced at $P$ by all elements contained in $J$ are approximated using multipole expansion coefficients, else if $J$  is leaf (hached boxes in Figure \ref{schematicFMMNLOGN}), these vectors are calculated by direct summation, else, we go to the next level and repeat the same procedure for $J$ children (such as blue boxes in Figure \ref{schematicFMMNLOGN}).
\end{itemize}

\end{itemize}

\end{enumerate}

\section{Introducing $n_F$ and $n_T$ through cost analysis}\label{sec:costAnalysis}

The speed of the adaptive scheme is function of $p$, $n_D$, and $l_m$, where $l_m$ is the maximum level reached upon tree division. As $p$ or  $n_D$ increases, the speed of the adaptive scheme will always decrease as we are targeting a more accurate approximations. However, for a given $p$ and  $n_D$, as $l_m$ increases the speed of the adaptive scheme increases until it reaches a critical level above which it would be faster and more accurate to perform direct summation rather than proceding to higher level.\\

Let $a$ be the number of operation needed to calculate the direct interaction at a target $i$ due to a point source $j$, $b$ the number of operation needed to calculate the interaction due to a set of sources located inside a box $J$ at a given target $i$  using multipole expansion coefficients, $c$ the number of operation needed to get local expansion coeficients from multipole expansion coefficients, and $d$ the number of operation needed to translate local expansion coeficients of a certain box to one of its children.\\

For a given $p$ and $n_D$, let us consider a target box $I_l$ at level $l$ that contains $n$ Elements. Next we present a criteria based on the number of elements contained in a target box $I_l$ that will help us to judge whether it is faster to go further one level down. We denote $n_T$ the critical number of element in box $I_l$ below which dividing this box into eight children and evaluating their influence using multipole expansion coefficients rather than direct summation will lower the speed of the adaptive fast solver, and $n_F$ the critical number of element in box $I_l$ below which dividing this box into eight children and evaluating their influence using local expansion coefficients rather than direct summation will lower the speed of the adaptive fast solver.\\

\subsection{Evaluating $n_T$ through cost analysis}

For a given target element $i$ located inside a source box $I_l$ at level $l$, the number of operation needed to calculate the direct interaction between all elements within neighbor boxes  of box $I_l$ and element $i$ is equal to $8(n_D+0.5)^3 na$ since we have $8(n_D+0.5)^3 $ neighbor boxes containing $n$ Elements each.\\

Now let us consider the same target element $i$ contained within a box $I_{l+1}$ at level $l+1$ where box $I_{l+1}$ is a child of box $I_l$. The number of operation needed to calculate the direct interaction between all elements within neighbor boxes of box $I_{l+1}$ and element $i$ is equal to  $(n_D+0.5)^3 na$ since we have $8(n_D+0.5)^3 $ neighbor boxes containing $\frac{n}{8}$ elements each. The number of operation needed to calculate the interaction between all elements within the interaction list of box $I_{l+1}$ and element $i$ is equal to $7\times 8(n_D+0.5)^3 b$ since we have $7\times 8(n_D+0.5)^3 $ boxes which are children to box $I_l$  and are not neighbor to box $I_{l+1}$.\\

The critical number $n_T$ can be found when we have the same amount of operations in both cases, that is:

\begin{equation}
8(n_D+0.5)^3 n_T a =(n_D+0.5)^3 n_T a+7(n_D+0.5)^3 \times 8\times b
\end{equation}

\begin{equation}
n_T=\frac{8b}{a}
\label{criticaltreecode}
\end{equation}

We conclude that if a target box $I_l$ at level $l$ contains $n$ elements, it would be beneficial to go to level $l+1$ and calculate the velocity and vortex stretching vectors induced by all sources in $I_l$ neighours using multipole expansions if and only if $n>n_T$. This critical number of elements is independent of $n_D$, and since $b$ is of the order of $p^2$, we can conclude that

\begin{equation}
n_T = xp^2+yp+z
\end{equation}

Where $x$, $y$, and $z$ are three constants determined by the computer system, language, implementation, core function, etc. These constants also depends on the field that need to be evaluated at every step.\\

Now, if we consider a source box $J$ that contains $n$ elements, the number of operation needed to calculate the direct interaction between $J$ and any target elements $i$ outside $\mathcal{N}_{n_D}(J)$ is equal to $na$. However, since $i$ is outside $\mathcal{N}_{n_D}(J)$, the interaction between $J$ and $i$ can be performed using one Multipole expansion. We conclude that if a source box $J$  contains $n$ elements, it would be beneficial to calculate the velocity and vortex stretching vectors induced by all sources in $J$ at any target element outside $\mathcal{N}_{n_D}(J)$ using multipole expansions rather than direct summation if and only if $n > \frac{b}{a}=\frac{n_T}{8}$.

\subsection{Evaluating $n_F$ through cost analysis}

For a given target element $i$ located inside a box $I_l$, the number of operation needed to calculate the direct interaction between all elements within neighbor boxes  of box $I_l$ and  element $j$ is equal to $8(n_D+0.5)^3 na$ since we have $8(n_D+0.5)^3 $ neighbor boxes containing $n$ elements each. Remembering that we have $n$ Elements in box $I_l$,  the number of operation needed to calculate the direct interaction between all source elements contained in neighbor boxes  of $I_l$ and all target elements contained within box $J_l$ is equal to $8(n_D+0.5)^3 n^2 a$\\

Now let us consider the same target element $i$ contained within a box $I_{l+1}$ at level $l+1$ where box $I_{l+1}$  is box a child of $I_l$ . The number of operation needed to calculate the direct interaction between all the elements contained in neighbor boxes of box $I_{l+1}$  and the element $j$ is equal to  $(n_D+0.5)^3 na$ since we have $8(n_D+0.5)^3 $ neighbor boxs containing $\frac{n}{8}$ elements each. Remembering that we have $n$ Elements in box $I_l$,  the number of operation needed to calculate the direct interaction between all source elements contained in neighbor boxs  of box $I_{l+1}$  and all target elements within box $I_{l+1}$  is equal to $(n_D+0.5)^3 n^2 a$\\

Furthermore, if we decided to go one level down, we will have additional cost as a consequence of the conversion and translation operations. The number of operation needed to perform the conversing operations is equal to $8\times 7\times 8(n_D+0.5)^3 c $ since we have $8$ children boxes and for each box  we have $7\times 8(n_D+0.5)^3  $ boxes which are children of box $I_{l+1}$  parent and are not neighbor to box $I_{l+1}$ . In addition, the number of operation needed to perform all the translation operations is equal to $8d$ \\

The critical number $n_F$ can be found when we have the same amount of operations in both cases, that is:

\begin{equation}
8(n_D+0.5)^3 n_F^2 a = (n_D+0.5)^3 n_F^2 a+8\times 7 \times 8(n_D+0.5)^3 c+8d
\end{equation}

\begin{equation}
n_F^2 = 64\frac{c}{a}+\frac{8}{7(n_D+0.5)^3}\frac{d}{a}
\end{equation}

\begin{equation}
n_F^2 = 64\left(\frac{c}{a}+\frac{1}{n_I}\frac{d}{a}\right)
\label{criticalFMM}
\end{equation}

Where $n_I$ is the maximum number of box in the interaction list. We conclude that if a box  at level $l$ contains $n$ elements, it would be beneficial to go to level $l+1$ and calculate the velocity and vortex stretching vectors using cell-cell interaction if and only if $n>n_F $. \\

This critical number is independent of $n_D$, and since $c$ and $d$ are of the order of $p^4$, we can conclude that:

\begin{equation}
n_F^2 = xp^4+yp^3+zp^2 +up+w
\end{equation}

Where $x, y, z, u, w $ are five constats determined by the computer system,  language, implementation, etc. These constants also depends on the field that need to be evaluated at every step. \\

Now, if we consider a source box $J$ that contains $N_S$ elements, the number of operation needed to calculate the direct interaction between $J$ and any target box $I$ outside $\mathcal{N}_{n_D}(J)$ containing $N_T$ elements is equal to $N_S*N_T*a$. However, since $I$ is outside $\mathcal{N}_{n_D}(J)$, the interaction between $J$ and $I$ can be performed using cell to cell interation with $c$ operation, and can also be performed using cell to particles with $N_T*b$ operations. We conclude that if a source box $J$ contains $N_S$ elements and a target box contains $N_T$ elements, it would be beneficial to calculate the velocity and vortex stretching vectors induced by all sources in $J$ at any target element inside $J$ with $I$ outside $\mathcal{N}_{n_D}(J)$ using cell-cell interation rather than direct summation or cell to particle interaction if and only if $N_S*N_T > \frac{c}{a}\simeq \frac{n_F^2}{64}$ and $N_T >\frac{c}{b} \simeq \frac{n_F^2}{8n_T}$.

\section{Error analysis of the vector potential for the adaptive scheme with regularized vortices}\label{sec:errorAnalysis}

\noindent In the case of regularized vortex method where the vorticity field is approximated using equation (\ref{eq:regularizedVorticityRepresentation}), the use of equation \ref{eq:multipoleExpansionApproximation} and  \ref{eq:localExpansion} to approximate the far-field vector potential will induce an error that has three major components, namely:
\begin{enumerate}
\item Error resulting from truncating Multipole Expansions at some order $p$, $\vec{E}_{M,\vec{\psi}}$: This error is related to the series expansion used to represent the kernel $\frac{1}{r}$ at large distance. The Multipole expansion error is related to the order of expansion $p$ and the Multipole Acceptance Criteria. $\parallel\vec{E}_{M,\vec{\psi}}\parallel$ decreases exponentially as $p$ increases and will eventually converge to zero as $p\rightarrow \infty$. However, increasing $p$ will have an high negative impact on the adaptive FM speed.
\item Error resulting from truncating Local Expansions at some order $p$, $\vec{E}_{L,\vec{\psi}}$: This error is also related to the series expansion used to represent the kernel $\frac{1}{r}$ at large distance and decreases exponentially as $p$ increases.
\item Error resulting from approximating the Biot-Savart kernel by a $\frac{1}{r}$ kernel, $\vec{E}_{\sigma,\vec{\psi}}$: This error arose due to the fact that $G_{\sigma}$ deviates from $\frac{1}{4\pi r}$. This error mainly depend on the size of the smallest boxes (at the deepest level of the tree). In order to keep $\parallel \vec{E}_{\sigma,\vec{\psi}} \parallel$ small, we must keep the leaf boxes width above a minimum multiple of the core function smoothing raduis. In three dimensional flows, this will pose a considerable computational overload on the FM schemes, since it will dramatically increase the particle to particle interaction.
\end{enumerate}

\noindent Now let us consider a target element $i$ within a leaf box $I_l^{*}$. The adaptive scheme can perform cell-cell, cell-particle, and particle-particle interactions. $\cal M$ is the set of all boxes $J$ at different levels that interact with element $i$ via cell-particle interaction, that is, the vector potential induced at $i$ by all vorticies contained in $J$ is calculated using equation \ref{eq:multipoleExpansionApproximation}. $\cal L$ is the set of all boxes at different levels that interact with cell $I_l$ via cell-cell interaction, that is, the vector potential induced at $i$ by all vorticies contained in $J$ is calculated using equation \ref{eq:localExpansion}. The total error can be expressed as:

\begin{align}
\lVert\vec{E}_{t,\vec{\psi}} \rVert & =\lVert \vec{E}_{\sigma,\vec{\psi}}+\vec{E}_{M,\vec{\psi}}+\vec{E}_{L,\vec{\psi}} \rVert
\end{align}

with

\begin{equation}
E_{\sigma,\vec{\psi}} \le \Gamma_{\cal M \cal L} \beta_{\sigma}(r^*)
\end{equation}

\begin{equation}
\lVert\vec{E}_{M,\vec{\psi}}\rVert \le  \frac{1}{4\pi}\frac{\Gamma_{\cal M}}{W_0}\frac{2^{l_m}}{n_D-\frac{\sqrt{3}-1}{2}}\left(\frac{\sqrt{3}}{2 n_D+1}\right)^{p+1}
\label{eq:streamlineTruncationError}
\end{equation}

\begin{equation}
\lVert\vec{E}_{L,\vec{\psi}}\rVert \leq \frac{1}{4\pi} \frac{\Gamma_{\cal L}}{W_{0}} \frac{2^{l_m}}{n_D+1-\sqrt{3}}\left(\frac{1}{\frac{2\sqrt{3}}{3}(n_D+1)-1}\right)^{p+1}
\end{equation}
\\

\noindent where $r^* =(n_D+0.5)\frac{W_0}{2^{l_m}}$, $\beta (r) =\abs{G(r)-\frac{1}{4\pi r} } $, $\beta_{\sigma}(r) = \frac{1}{\sigma} \beta \left( \frac{r}{\sigma}\right)$, $W_0$ is the width of the root box, and $\mathcal{ML}=\mathcal{M} \cup \mathcal{L}$. Furthermore, for any set of boxes $\cal S$, we have:

\begin{equation}
\Gamma_{\cal S}=\sum_{j \in \cal S}\lVert\vec{\alpha}_j\rVert
\end{equation}

\noindent Let us consider a specific core function with core radius equal to $\sigma$. For a given tree depth $l_m$, as $p$ increases $\lVert \vec{E}_{\sigma,\vec{\psi}} \lVert $ remains constant while $\lVert \vec{E}_{M,\vec{\psi}} \lVert $ and $\lVert \vec{E}_{L,\vec{\psi}} \lVert $decrease continuously. Thus the total error converge to $ \lVert \vec{E}_{\sigma,\vec{\psi}} \lVert$ as $p \rightarrow \infty$ . So, for each tree depth $l_m$, there is a order of expansion $p^{m}$ above which increasing $p$ won't considerably improve the accuracy since the core function (regularization) error would be dominant.\\

\section{ Error analysis of the velocity vector for adaptive FM scheme with regularized vortices}\label{sec:errorAnalysis}

The velocity vector is also expressed as a truncation of spherical harmonics. However it has different expressions than those obtained for vector potential, and thus it has different error bounds. The error in the velocity vector can also be decomposed into three major components, namely, the Multipole truncation error $E_{M,\vec{u}}$, the Local truncation error $E_{L,\vec{u}}$ and the regularizing error $E_{\sigma,\vec{u}}$. The total error can be expressed as:

\begin{align}
\lVert\vec{E}_{t,\vec{u}} \rVert & =\lVert \vec{E}_{\sigma,\vec{u}}+\vec{E}_{M,\vec{u}} + \vec{E}_{L,\vec{u}}\rVert
\end{align}

with 

\begin{equation}
\lVert\vec{E}_{\sigma,\vec{u}} \rVert\leq  \Gamma_{\cal M \cal L} \kappa_{\sigma}(r^*)
\end{equation}

\begin{equation}
\lVert\vec{E}_{M,\vec{u}} \rVert \le  \frac{1}{4\pi}\frac{\Gamma_{\cal M}}{W_0^2}\frac{4^{l_m}}{\left(n_D-\frac{\sqrt{3}-1}{2}\right)^2}\left(\frac{\sqrt{3}}{2 n_D+1}\right)^{p+1} \left[p+2-\frac{\sqrt{3}}{2n_D+1}(p+1)\right]
\end{equation}

\begin{equation}
\lVert\vec{E}_{L,\vec{u}} \rVert\le \frac{1}{4\pi} \frac{\Gamma_{\cal L}}{W_0^2} \frac{ 4^{l_m}}{\left(n_D+1-\sqrt{3}\right)^2}\left( \frac{1}{\frac{2\sqrt{3}}{3}(n_D+1)-1}\right)^{p+1}\left[p+2-\left( \frac{1}{\frac{2\sqrt{3}}{3}(n_D+1)-1}\right) p \right]
\end{equation}
\\

\noindent where $\kappa (r)= \frac{1}{4\pi r^2}\arrowvert 1-4\pi q(r) \arrowvert $ and $\kappa_{\sigma}(r)=\frac{1}{\sigma^2}\kappa (\frac{r}{\sigma})$.\\

\noindent $E_{\sigma,\vec{u}} $ is function of the choice of the core function and its radius, and mainly depends on the maximum level $l^{m}$ reached upon tree division. For a given level of accuracy, there exist a critical level $l_c$ above which the regularization error would be higher than the maximum allowable error. Thus the tree division must stop whenever the tree level reaches $l_c$. On the other hand, for a constant $n_D$, the truncation errors, $E_{M,\vec{u}}$ and $E_{L,\vec{u}}$, mainly depend on the order of expansion $p$. For a given level of accuracy, there exist a critical level $p_c$ below which the truncation error would be higher than the maximum allowable error. For a given order of expansion $p_c$, the adaptive solver evaluates two critical numbers $n_T$ and $n_F$ and also uses them to limit the depth of the tree in order to balance near-field and far-field evaluations and optimze the speed of the fast solver. In fact, the division process will stop whenever the number of elements in a given box is less than $n_C = \min \{n_T,n_F \}$. Let $l^{*}$ be the tree level reached when we stop the division when number of elements in a given box is less than $n_C = \min \{n_T,n_F \}$ regardless of $l_c$. This would be the case when we use a Dirac delta function. If $l_c > l^{*}$ the fast solver will manage to balance the loads. However if  $l_c < l^{*}$ the tree structure would be too shallow and the near field evalutions dominates. In this case, the adpative solver behaves exacly the same as Greengard FMM since the division process is halted at a given level.

\section{Results and Discussion}

\subsection{Computational cost}
 
In the context of vortex methods, the adaptive fast multipole scheme developed throughout this paper uses both multipole and local spherical harmonic based expansions to approximate the far-field velocity and vortex stretching vectors. This scheme extends Greengard \cite{Greengard:1987kx} fast multipole algorithm by introducing a new mechanism to optimize the speed of  the fast multipole method. The introduced mechanism make use of two critical number  $n_T$ and $n_F$ developed in section \ref{sec:costAnalysis} to balance the far- field and near field evaluations for different architectures. Both $n_T$ and $n_F$ depends on the order of expansion $p$  and the scalar and vector fields (kernel) evaluated by the adaptive scheme.  For a given architecture and given order of expansion $p$, execution time of four different kernels is measured. This information is used to find the optimum values of $n_T$ and $n_F$ through equations \ref{criticaltreecode} and \ref{criticalFMM}. These values are used in the context of  \ON fast multipole to limit the depth of the tree structure and to balance the computational cost of particle to particle, cell to particle, and cell to cell interaction as described in section \ref{sec:fastN}.\\

\noindent Using singular vorticies, we have performed several simulation for different order of expansion $p$ ranging from $2$ to $20$. If the objective were to optimize the cost of  the velocity and vortex stretching vectors calculation, we found out that  $n_T = 4.802p^2 +16.74p+63.52$ and $n_F = 4.167p^2 +6.689p+13.91$ would maximize the speed of the adaptive fast multipole scheme. If the objective were to optimize the cost of  the velocity calculation, we found out that  $n_T = 3.112p^2 +13.35p+19.11$  and $n_F = 5.897p^2 +9.468p+19.69$ would  then maximize the speed of the adaptive fast multipole scheme. \\

\begin{figure}[!htb]
\begin{center}
\includegraphics[width=6in]{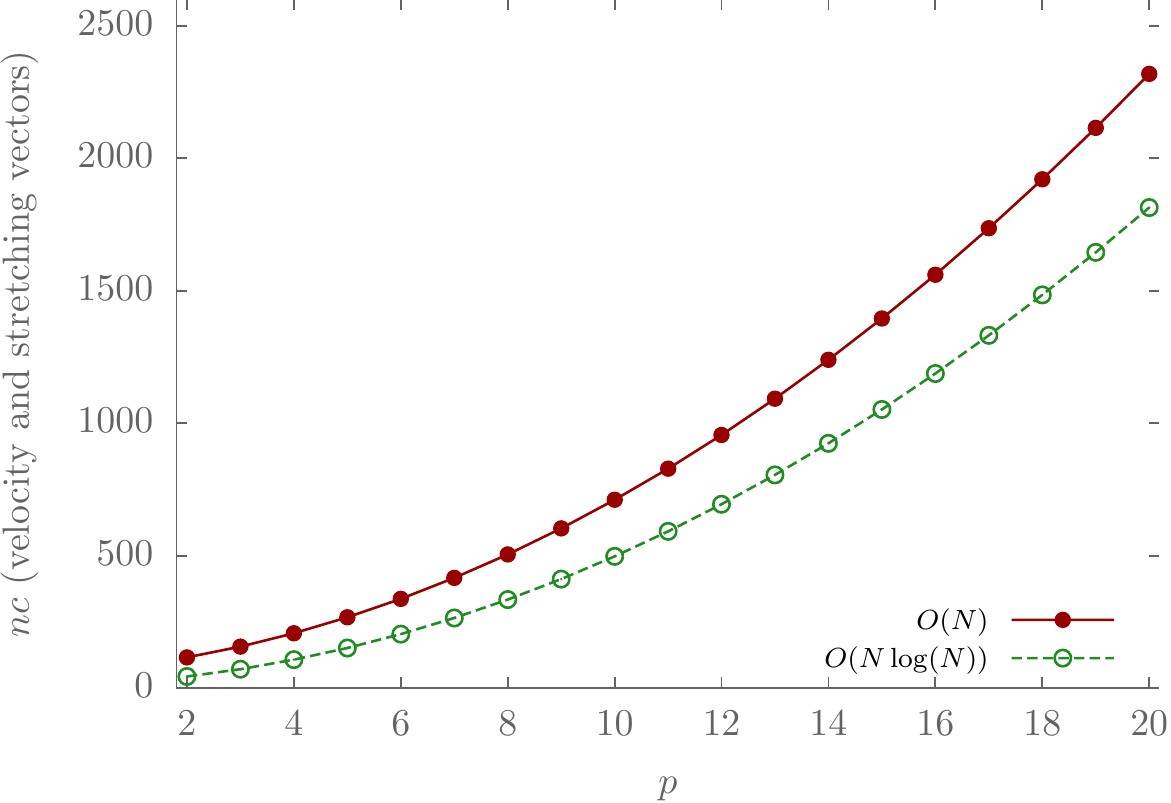}
\caption{Optimum $n_T$ and $n_F$ used when calculating both velocity and stretching vectors vesus $p$}
\label{CriticalBoth}
\end{center}
\end{figure}

\begin{figure}[!htb]
\begin{center}
\includegraphics[width=6in]{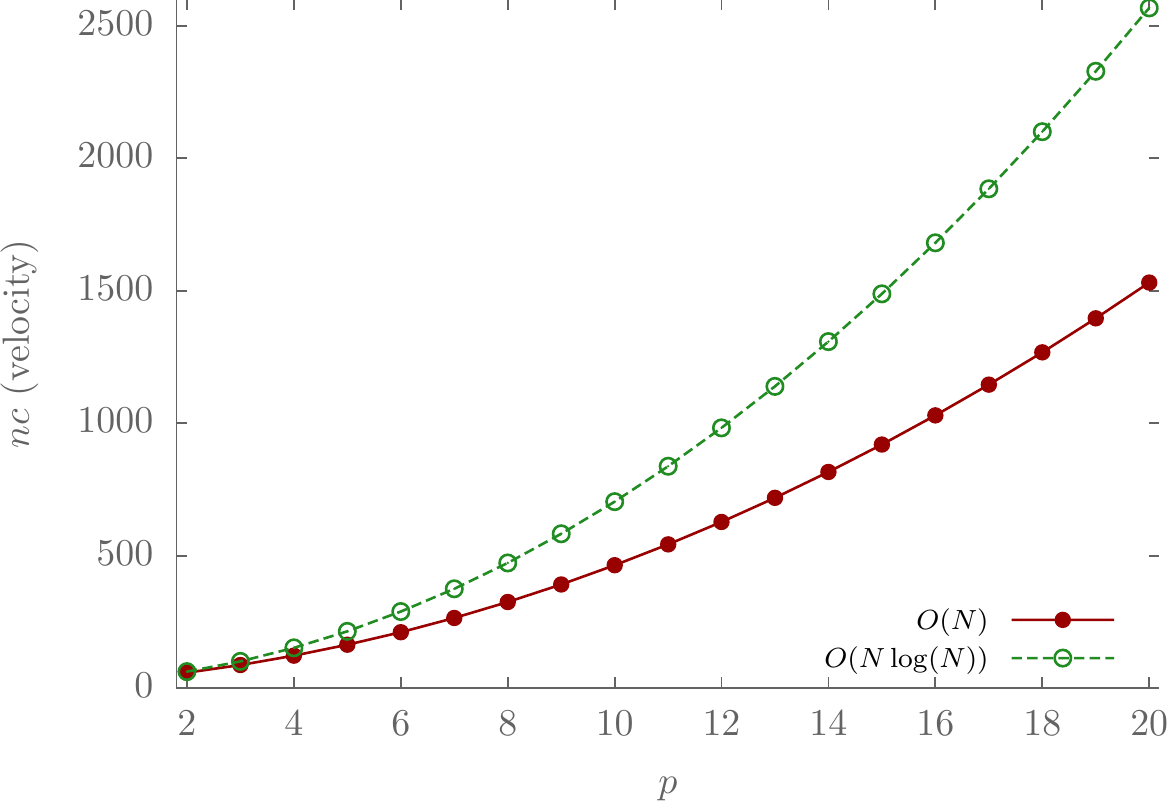}
\caption{Optimum $n_T$ and $n_F$ used when calculating only the velocity vector vesus $p$}
\label{CriticalVelocity}
\end{center}
\end{figure}

\noindent The $p$-dependence of the critical numbers $n_T$ and $n_F$ is shown in figures \ref{CriticalBoth} \ref{CriticalVelocity}. It is clear that, for all cases, as $p$ increases both $n_T$  and $n_F$ increase.  This is due to the fact that it is more expensive to calculate cell to particle and cell to cell interaction for larger $p$. Figure \ref{CriticalVelocity} shows the critical numbers used when only the velocity vector is calculated by the fast scheme. For this case, $n_F$ takes larger values than $n_T$ . Thus further cost reduction can be achieved by reaching higher level and by using cell to particle interaction at these levels to approximate the far-field velocity. At these levels, local expansion coefficients are not calculated since cell to cell interaction would increase the overall computational cost. However when both velocity and vortex stretching vectors are evaluated by the fast scheme,  $n_F$ takes on smaller values than $n_T$ for any order of expansion $p$. Thus local expansion coefficients are calculated for all cells at leaf level and the use of cell to particle interaction won't present much advantages. In this case, the adaptive scheme will basically calibrate the balance between particle to particle and cell to cell interaction to obtain the optimum speed.\\

\noindent The fact that $n_F$ becomes smaller than $n_T$ when both velocity and stretching vectors are calculated can be logically justified. In fact, cell to cell computational effort is mainly due to conversion and translation operations which are used to find local expansions coefficients at the center of each leaf cell. Once obtained, these coefficients are used to approximate the velocity,  vortex stretching, and any higher order derivative of the velocity vector field. So basically, this is a fixed cost that must be paid whether we want to calculate the velocity vector alone or both velocity and vortex stretching vectors and it is by far the most time consuming part in the cell- cell evaluations. Thus calculating the vortex stretching vector, in addition to velocity vector, would not impact cell to cell computational load since local expansion coefficients must be calculated either way. However, calculating the vortex stretching, in addition to the velocity, would dramatically increase the particle to particle load and would also increase cell to particle computational effort. Thus, to achieve the optimum speed, $n_F$ must decrease in order to balance particle to particle and cell to cell interations. On the other hand, the decrease of $n_T$ is rather limited because both particle to particle and cell to particle loads are increasing.\\

\noindent The values of $n_T$ and $n_F$ also depend on the regularization function used to evaluate the near filed velocity and vortex stretching via particle to particle interaction. In fact, compared to the optimum values obtained for the singular Dirac Delta function case, $n_T$ must be divided by $R$ and $n_F$  must be divided by $\sqrt{R}$, where $R = \frac{a_\sigma}{a_\delta}$.  $a_\delta$ and $a_\sigma$ are the number of operations needed to calculate the direct interaction for the case of singular and regularized vortices respectively. Thus, the regularization function has higher effect on $n_T$ than $n_F$.\\

To investigate the performance of the adaptive scheme, we carried out several simulations where both velocity and vortex stretching vectors are calculated simultaneously.  A vortex ring with a unit radius and a core radius equal to $0.1$ is considered. The initial vorticity within the core of the ring , taken as a second-order Gaussian distribution,  is discretized using $N$ singular vortex elements where $N$ ranges between $10^4$ and $20^6$. Throughout all the simulations, $n_D$  is set to $1$. This is equivalent to a typical fast multipole with a $3 \times 3 \times 3$ neighbor list. The adaptive scheme  performs particle to particle , cell-to particle, and cell to cell interaction. A mechanism to choose among these interactions is based upon the critical numbers defined in section \ref{sec:costAnalysis}. 

\begin{figure}[!htb]
\begin{center}
\includegraphics[width=6in]{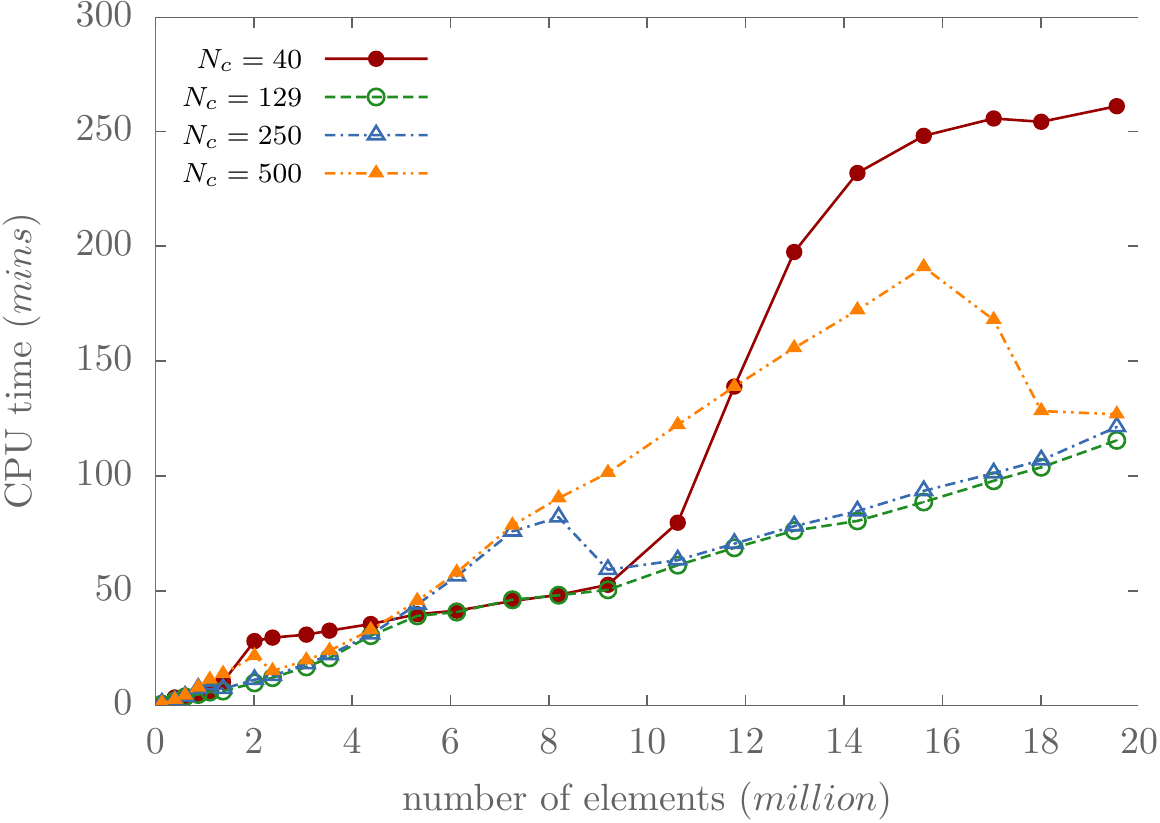}
\caption{Measured simulation time versus the number of elements for different $n_F$ using an order of expansion $p=5$}
\label{AdaptiveP5}
\end{center}
\end{figure}

The timings on a single CPU core for the hybrid method are shown in Figure \ref{AdaptiveP5}.  The order of expansions is $p = 5$ and the simulations are conducted for a wide range of number of elements and for the following values of critical number $n_F = 40, 129, 250, 500$,  where $n_F = 129$ is the optimum value chosen by the scheme that should optimize the speed of the adaptive scheme for $p=5$ by balancing near-field and far-field evaluations.
As it can be seen from figure \ref{AdaptiveP5}, and for all values of the number of elements, the optimum value chosen by the scheme ,$n_F=129$, gives the best performance. In fact for $n_F =40, 250, 500$, the adaptive scheme  suffers from load imbalance between the near-field and far-field evaluations. Particle to particle interaction load is dominant for $n_D=250$ and $500$, while cell to cell interaction are dominant for the case od $n_D=40$ since we have to calculate local expansions for cells at a deep tree structure.

Figure \ref{AdaptiveP5} also shows that as  we deviate from the optimum value of $n_F$, simulation time begin to oscillate above the  time measured for the optimum  case of $n_F=129$. These oscillations takes higher peak value  and smaller frequency as we move away from the optimal value. Load imbalance  caused by improper value of $n_F$ might double the simulation time.

\begin{figure}[!htb]
\begin{center}
\includegraphics[width=6in]{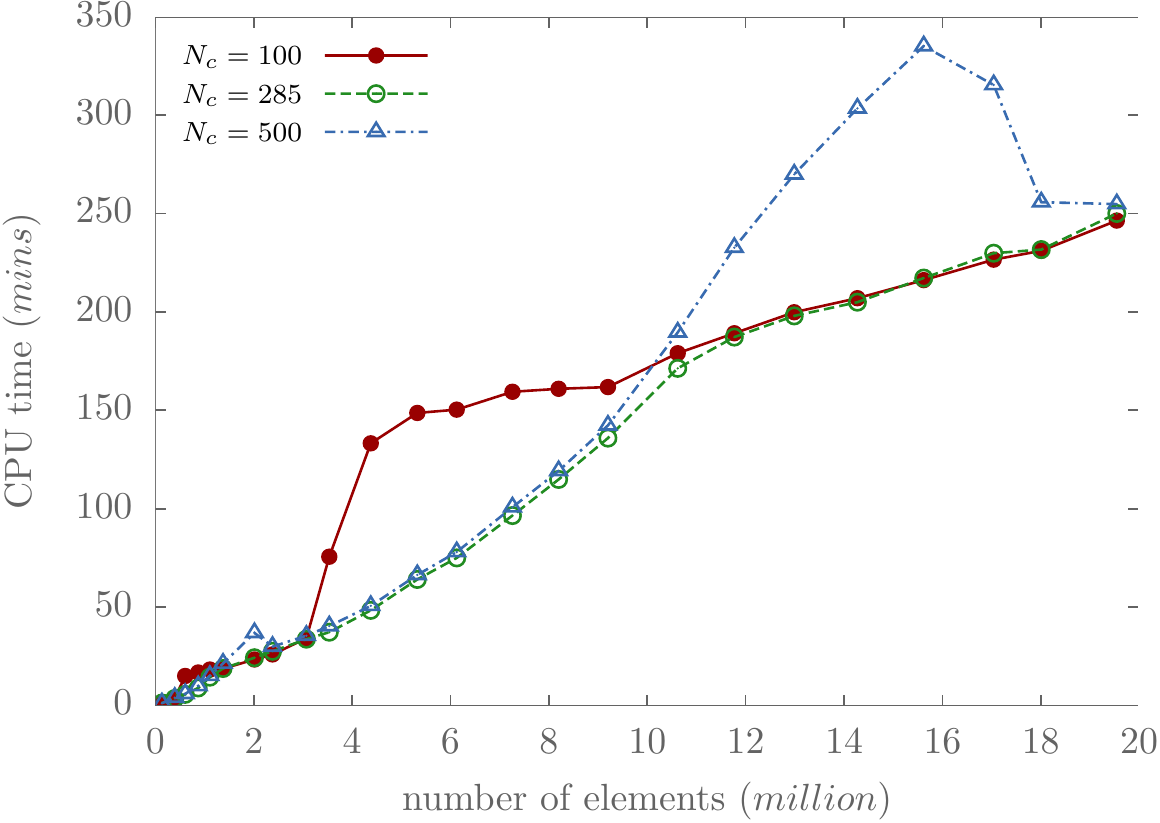}
\caption{Measured simulation time versus the number of elements for different $n_F$ using an order of expansion $p=8$}
\label{AdaptiveP8}
\end{center}
\end{figure}

Same simulations were conducted for an order of expansion $p=8$. In this case the optimum critical value is $n_F=285$. The timing measured in this case is compared with that obtained for $n_F=100$, and $500$. Figure \ref {AdaptiveP8} shows that optimum value of $n_F$ gives the best performance and the results are similar to those obtained for $p=5$

We conclude that the optimum case has major advantages. By limiting the depth of the tree structure and balancing near-field and far-field evaluation, the adaptive scheme optimizes the speed of the fast multipole method for a given order of expansion.  This scheme is flexible and has auto-tuning capabilities on heterogeneous architectures, and can run on any machine without changing anything. Moreover, the current method will automatically choose the critical numbers and  alleviate the user from this burden. In addition, the adaptive scheme scale as $\mathcal{O}(N^{1.045})$ and eliminates the number of elements dependence of the method.

\subsection{Error analysis}

The proper evaluation of $n_T$ and $n_F$ would optimize the speed of the adaptive solver for a given order of expansion $p$ while using a Dirac Delta function. In this case, the accuracy of the fast summation is completely determined by the order of expansion. However, the use of a regularization core function instead of dirac function will induce an additional error, which depends on the used core function and the neighborhood dimension. For a given level of accuracy, there always exist a critical level $l_c$ that we should not cross over.\\

\begin{figure}[!htb]
\begin{center}
\includegraphics[width=6in]{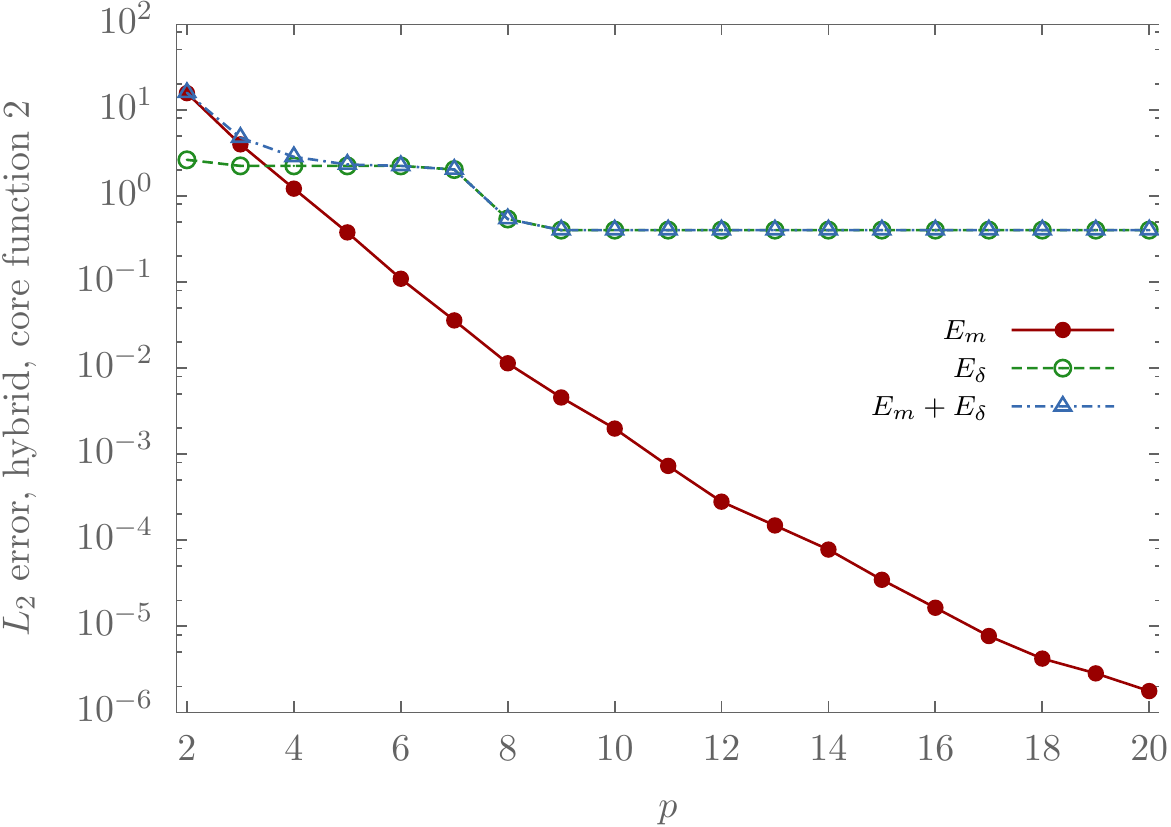}
\caption{$L_2$ norm for the truncation and regularization error verus $p$ using a second order algebriac core function in the context of adaptive scheme}
\label{NCore2}
\end{center}
\end{figure}

\begin{figure}[!htb]
\begin{center}
\includegraphics[width=6in]{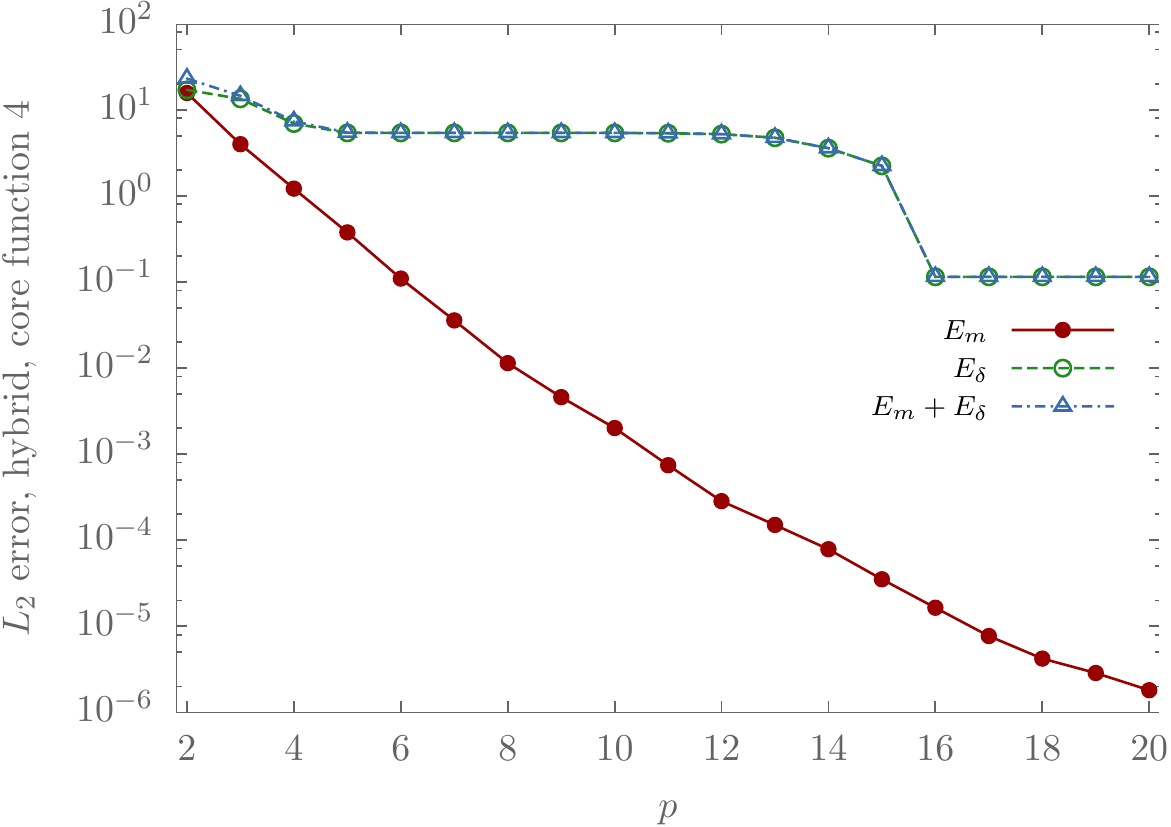}
\caption{$L_2$ norm for the truncation and regularization error verus $p$ using a second order Gaussian core function in the context of adaptive scheme}
\label{NCore4}
\end{center}
\end{figure}

\begin{figure}[!htb]
\begin{center}
\includegraphics[width=6in]{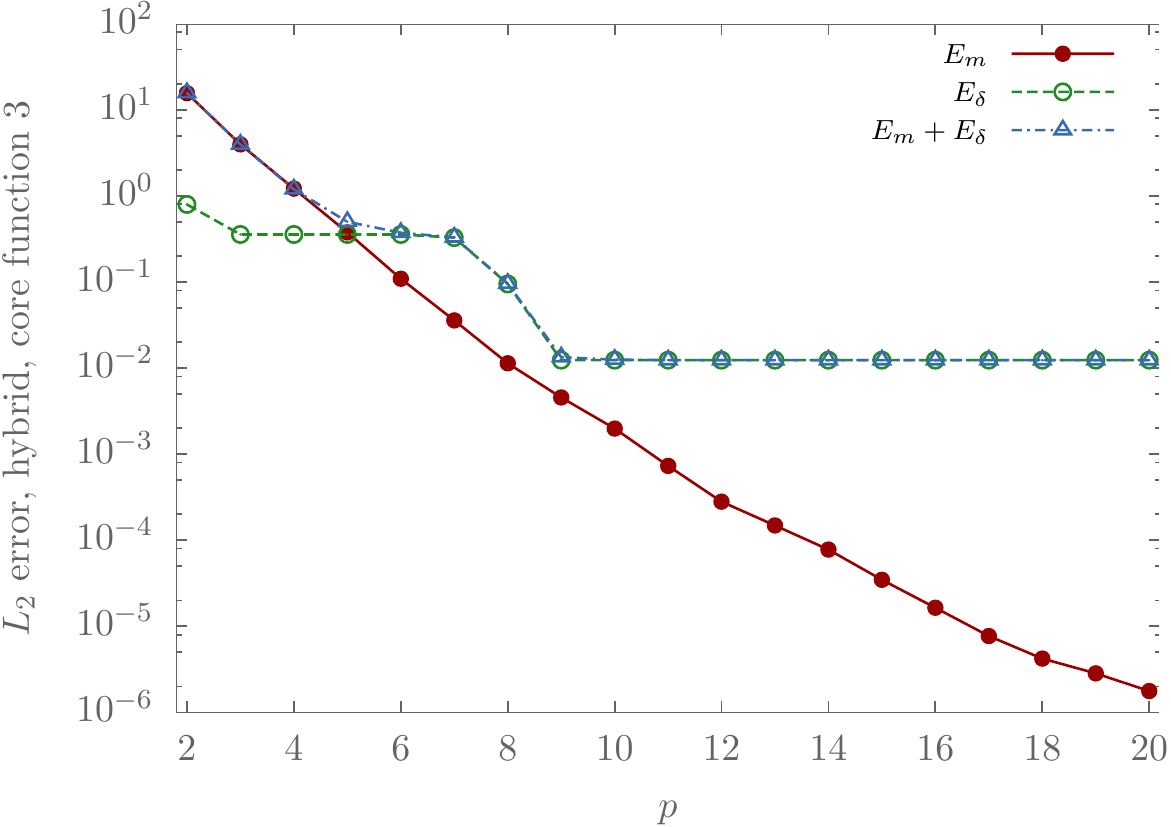}
\caption{$L_2$ norm for the truncation and regularization error verus $p$ using a fouth order algebriac core function in the context of adaptive scheme}
\label{NCore3}
\end{center}
\end{figure}

\noindent To investigate the accuracy of the adaptive scheme, we have conducted several simulation for different order of expansion $p$ and using different core function and we have reported the $L_2$ norm of the truncation and regularization error.  A vortex ring with similar geometric parameter to that described earlier is descritized using $N=1108880$ elements. Figures \ref{NCore2} \ref{NCore4} \ref{NCore3} show  the $L_2$ error norm versus $p$ for a second order algebraic core function, second order Gaussian core function, and fourth order algebraic core function respectively obtained for the adaptive solver. As it can be seen in all cases, while the  truncation error always decreases exponentially, the regularization error and thus the total error decreases in steps from one stable value to another. These error values correspond to regularization error calculated when we force the tree division to stop at a certain level. In fact, the regularization error depends mainly on the core function and on the distance to neighbor cells which is related to the tree depth $l_m$. For a given core function, increasing $p$ will lead to an increase in $n_T$ and $n_F$ which might lead to decrease the tree depth. This mainly explain the step behaviour of the regularization and total error.

\begin{figure}[!htb]
\begin{center}
\includegraphics[width=6in]{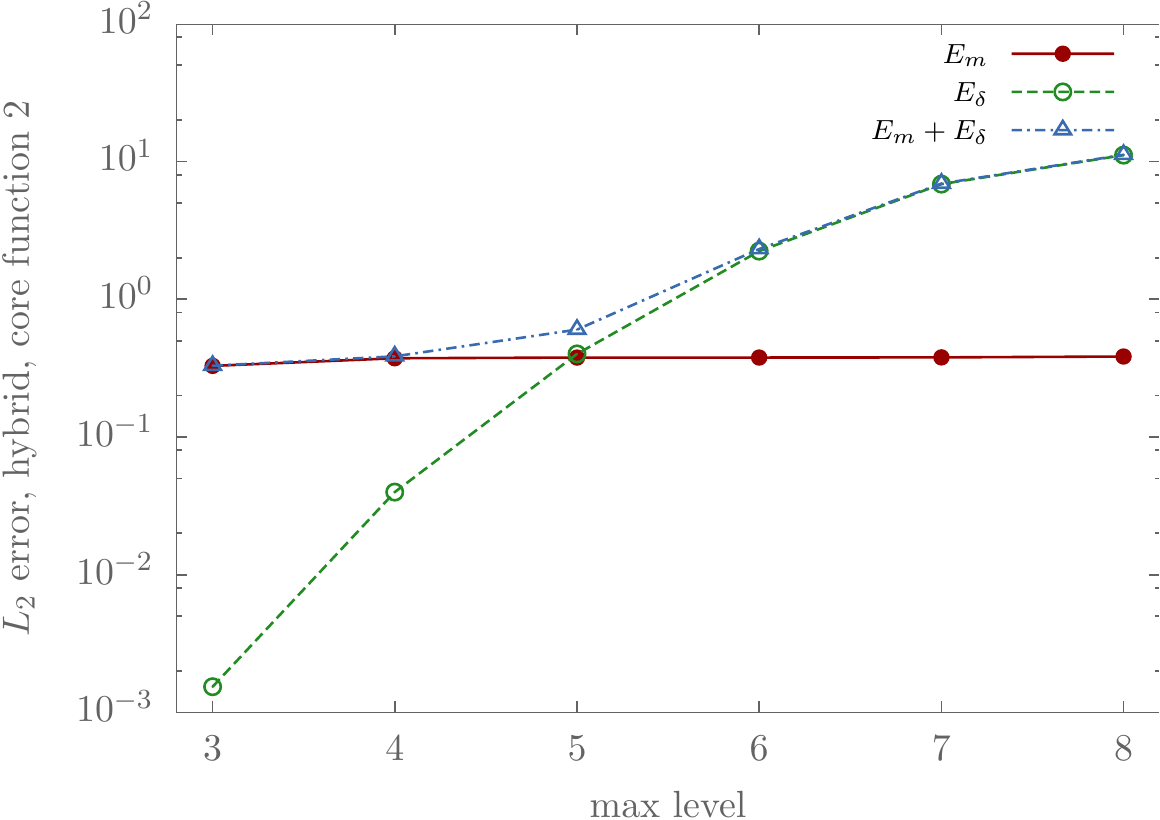}
\caption{$L_2$ norm for the truncation and regularization errors verus $l_m$ for a fixed order of expansion $p=5$ using a second order algebraic core function in the adaptive scheme}
\label{NLeaf2}
\end{center}
\end{figure}

\begin{figure}[!htb]
\begin{center}
\includegraphics[width=6in]{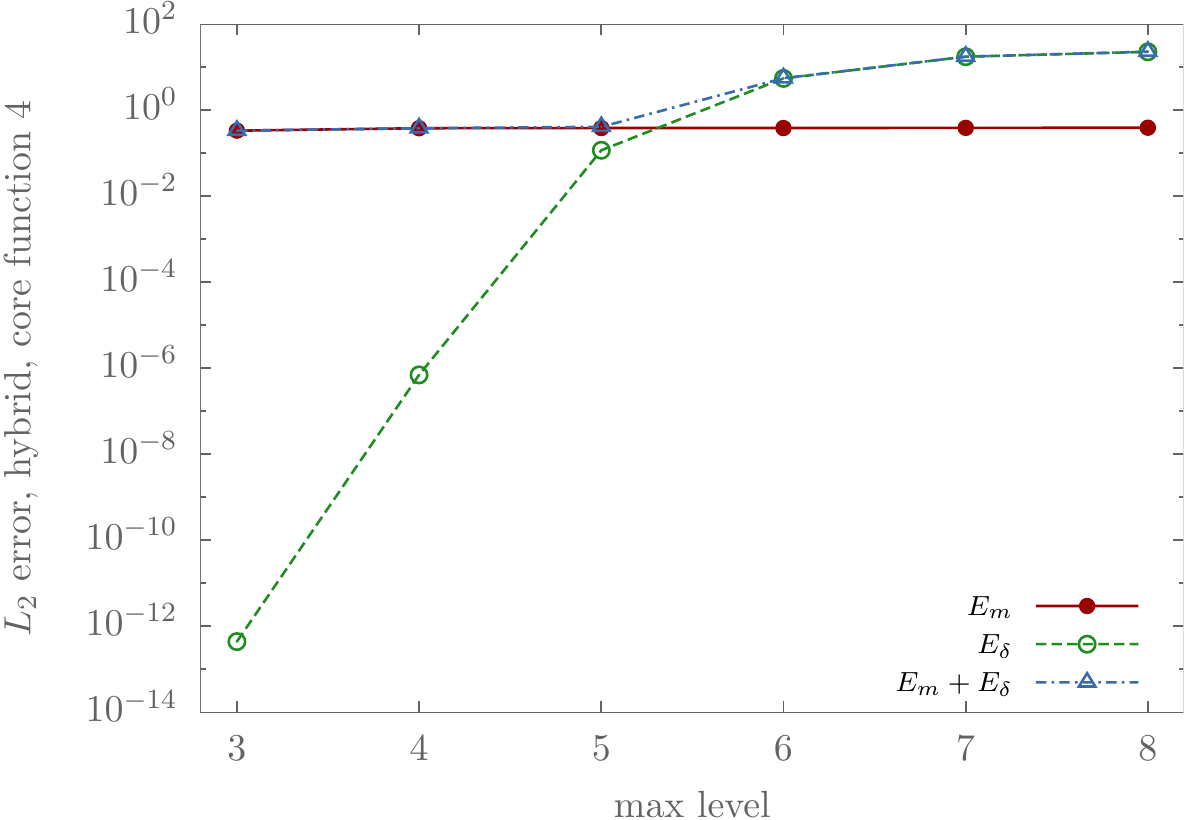}
\caption{$L_2$ norm for the truncation and regularization errors verus $l_m$ for a fixed order of expansion $p=5$ using a second order Gaussian core function in the adaptive scheme}
\label{NLeaf4}
\end{center}
\end{figure}

\begin{figure}[!htb]
\begin{center}
\includegraphics[width=6in]{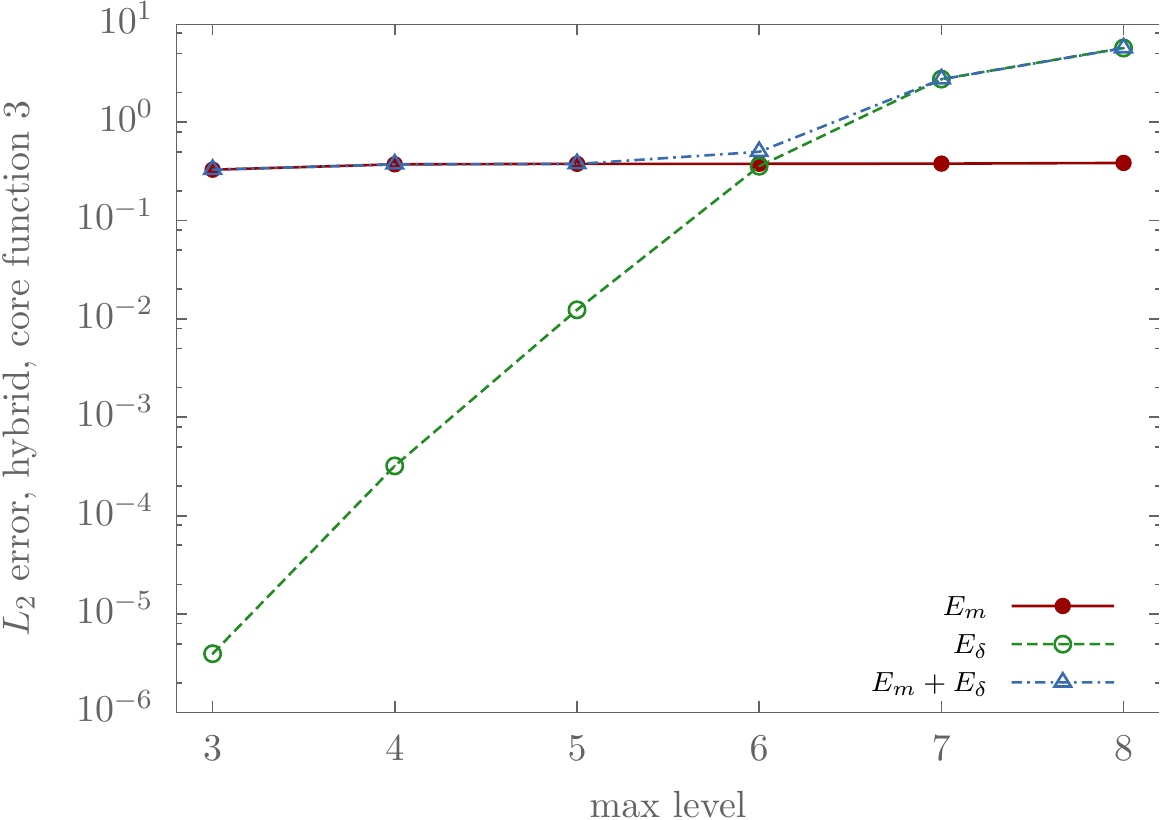}
\caption{$L_2$ norm for the truncation and regularization errors verus $l_m$ for a fixed order of expansion $p=5$ using a fourth order algebraic core function in the adaptive scheme}
\label{NLeaf3}
\end{center}
\end{figure}

Figure \ref{NLeaf2} \ref{NLeaf4} \ref{NLeaf3} show the regularization error obtained when we have force the division of the tree to stop at different levels for the three core functions for a fixed order od expansion $p=5$. It is clear that as the maximum level $l_m$ increases, the regularization error increases while the truncation error remains constant. The increase in the regularization error is because increasing the tree level $l^m$ will decrease the distance between neighboring cell and thus increase $\kappa_\sigma$.

From all the above, we conclude that, for a given level of accuracy, there exist a critical level $l_c$ above which the regularization error would be higher than the allowable error. For this critical level, there exist a critical order of expansion $p^c$ above which increasing $p$ will only increase the computational cost without improving the total error. The critical level $l_c$ and its associated critical order of expansion $p_c$ are first evaluated by evaluating the errors aon a representive sample of vortices. Then, the critical number of elements $n_T$ and $n_F$ are evaluated as a function pf $p_c$.  The division process stops whenever the box level $l$ is equal to the critical level $l_c$ or the number of elements in the box is less $n_C = \min \{n_T,n_F \}$, where $n_T$ and $n_F$ are defined in section \ref{sec:costAnalysis}.

\section{Conclusions}
\label{sec:conclusions}

We presented an error-controlled adaptive fast solver for approximating the velocity and vortex stretching vectors in the context of grid free three dimensional vortex methods. We introduced three critical numbers, $n_T, n_F$, and $l_c$.  These critical numbers are used to limit the tree depth and to balance near and far field evaluations in order to obtain the highest speed for any core function and any level of accuracy.  The adaptive scheme scale as $\mathcal{O}(N^{1.045})$ and eliminates the number of elements dependence of the method. This scheme is flexible and has auto-tuning capabilities on heterogeneous architectures, and can run on any machine without changing anything. Moreover, the current method will automatically choose the critical numbers and  alleviate the user from this burden.

\section{Acknowledgments}
\label{sec:ackn}

This work is supported by the University Research Board (URB) of the American University of Beirut.






\section{Bibliography}

\bibliographystyle{elsarticle-harv}
\bibliography{vortex}
\newpage

\section{Appendix A: Error bounds for the potential and velocity vectors}

\subsection{Error bounds for the potential vector induced by a singular source element}
Consider a singular source element $j$ of strength $\vec{\alpha}_j$ located at $Q=(\rho_j,\theta_j,\varphi_j)$. The potential vector induced by $j$ at a target element located at $P=(r,\theta,\phi)$ has the following expression:
\begin{equation}
\vec{\psi}_j(P)=\frac{\vec{\alpha}_j}{4\pi d_j},
\end{equation}
where $d_j=\parallel\vec{QP}\parallel$ is the distance between source and target elements. Let $\gamma$ be the angle between the vectors $P$ and $Q$. From the law of cosines, we have:
\begin{equation}
d_j^2=r^2+\rho_j^2-2r\rho_j cos\gamma
\end{equation}
From this relation, we may write
\begin{equation}
\vec{\psi}_j(P)=\frac{\vec{\alpha}_j}{4\pi r}\frac{1}{ \sqrt{1-2\left(\frac{\rho_j}{r}\right) cos\gamma +\left(\frac{\rho_j}{r}\right)^2}}
\end{equation}
Let $\mu_j =\frac{\rho_j}{r}$. For $\rho_j < r$, that is $\mu_j < 1$, if we expand $\vec{\psi}(P)$ in term of Legendre Polynomial we obtain:
\begin{equation}
\vec{\psi}_j(P)=\frac{\vec{\alpha}_j}{4\pi r}\sum_{n=0}^{\infty} \mu_j^n P_n(cos \gamma)
\end{equation}
The error resulting from truncating $\vec{\psi}_j(P)$ at some order $p$ is:
\begin{equation}
E_{\vec{\psi}_j}=\left|\left| \frac{\vec{\alpha}_j}{4\pi d_j} - \frac{\vec{\alpha}_j}{4\pi r} \sum_{n=0}^{p} \mu_j^n P_n(cos \gamma)\right|\right|=\left|\left| \frac{\vec{\alpha}_j}{4\pi r}\sum_{n=p+1}^{\infty} \mu_j^n P_n(cos \gamma)\right|\right|
\end{equation}
Knowing that $P_n(cos \gamma) \le 1$, and using the triangle inequality, we obtain an error bound for the multipole expansion 
\begin{equation}
E_{\vec{\psi}_j} \le \left|\left| \frac{\vec{\alpha}_j}{4\pi r}\sum_{n=p+1}^{\infty} \mu_j^n \right|\right|\le \frac{\parallel \vec{\alpha_j}\parallel}{4\pi r}\frac{\mu_j^{p+1}}{1-\mu_j}
\end{equation}
\begin{equation}
E_{\vec{\psi}_j} \le \frac{1}{4\pi} \frac{\parallel \vec{\alpha_j}\parallel}{(r-\rho_j)}\left(\frac{\rho_j}{r}\right)^{p+1}
\end{equation}

Similary, when $r<\rho_j$, we define $\mu_j =\frac{r}{\rho_j}$. The potential vector and the error bound for the local expansion would have the following expressions:
\begin{equation}
\vec{\psi}_j(P)=\frac{\vec{\alpha}_j}{4\pi d_j}=\frac{\vec{\alpha}_j}{4\pi \rho_j}\sum_{n=0}^{\infty} \mu_j^n P_n(cos \gamma)
\end{equation}
\begin{equation}
E_{\vec{\psi}_j} \le \frac{1}{4\pi} \frac{\parallel \vec{\alpha_j}\parallel}{(\rho_j-r)}\left(\frac{r}{\rho_j}\right)^{p+1}
\end{equation}

\subsection{Error bounds for the potential vector induced by a cluster of source elements}
Suppose that $s$ vortices with strength $(\vec{\alpha}_j, j=1...s)$ are located at the points $\vec{Q}_j = (\rho_j,\theta_j,\varphi_j)$ inside the sphere $D_Q$ of radius $a$ with center at $Q=(0,0,0)$, then at any $\vec{P}=(r,\theta,\varphi)$ with $r>a$, the potential vector field is given by:
\begin{equation}
\vec{\psi}= \sum_j \vec{\psi}_j = \sum_j \frac{\vec{\alpha}_j}{4\pi d_j} =\sum_j \frac{\vec{\alpha}_j}{4\pi r} \sum_{n=0}^{\infty} \mu_j^n P_n(cos \gamma)
\end{equation}
The error resulting from truncating $\vec{\psi}(P)$ at some order $p$ is:
\begin{equation}
E_{\vec{\psi}}=\left|\left| \sum_j \left( \frac{\vec{\alpha}_j}{4\pi d_j} - \frac{\vec{\alpha}_j}{4\pi r} \sum_{n=0}^{p} \mu_j^n P_n(cos \gamma)\right)\right|\right|
\end{equation}
Using the triangle inequality, and since $\mu_j \le \frac{a}{r}$, the multipole expansion error will be bounded by:
\begin{equation}
E_{\vec{\psi}} \le \frac{1}{4\pi} \frac{\Gamma_{D_Q}}{(r-a)}\left(\frac{a}{r}\right)^{p+1}
\end{equation}
where $\Gamma_{D_Q} = \sum_j \parallel \vec{\alpha_j}\parallel$\\

\noindent Now suppose that $s$ vortices with strength $(\vec{\alpha}_j, j=1...s)$ are located inside a cubic box $J_l$ at level $l$ with center at $Q=(0,0,0)$. The radius of the smallest sphere which encloses box $J_l$ is $a_{J_l} = \frac{\sqrt{3}}{2}W_{J_l}$ where $ W_{J_l}$ is the box width. Following the adaptive solver Scheme described in section \ref{}, the distance between the center of box $J_l$ and any point $P =(r,\theta,\varphi) $ located outside the neighborhood of box $J_l$ is $r_{J_l} \geq (n_D+0.5)W_{J_l}$. For any $p \geq 1$, the multipole expansion error will be bounded by:

\begin{equation}
\vec{E}_{M,\vec{\psi}} \leq \frac{1}{4\pi}\frac{\Gamma_{J_l}}{W_{J_l}}\frac{1}{n_D-\frac{\sqrt{3}-1}{2}} \left(\frac{\sqrt{3}}{2 n_D+1}\right)^{p+1}
\end{equation}
Noting that $W_{J_l}=W_0/2^{l}$, we get
\begin{equation}
E_{M,\vec{\psi}} \le  \frac{1}{4\pi}\frac{\Gamma_{J_l}}{W_0}\frac{2^{l}}{n_D-\frac{\sqrt{3}-1}{2}}\left(\frac{\sqrt{3}}{2 n_D+1}\right)^{p+1}
\label{eq:streamlineTruncationError}
\end{equation}
where $W_0$ is the width of the root box, $\Gamma_{J_l} = \sum_j \parallel \vec{\alpha_j}\parallel$ and the summation is done over all the source elements within box $J_l$.\\

Similary, suppose that $s$ vortices with strength $ (\vec{\alpha}_i, i=1...s)$ are located inside the sphere $D_Q$ of radius $a$ with center at $Q=(\rho,\alpha,\beta)$, and that $\rho = (c+1)a $ with $c >1 $, then for any target $P=(r,\theta,\phi)$  inside the sphere $D_0$ of radius $a$ centred at the Origin, the potential vector is given by
\begin{equation}
\vec{\psi}= \sum_j \vec{\psi}_j = \sum_j \frac{\vec{\alpha}_j}{4\pi d_j} =\sum_j \frac{\vec{\alpha}_j}{4\pi \rho_j} \sum_{n=0}^{\infty} \mu_j^n P_n(cos \gamma)
\end{equation}
The error resulting from truncating $\vec{\psi}(P)$ at some order $p$ is:
\begin{equation}
E_{\vec{\psi}}=\left|\left| \sum_j \left( \frac{\vec{\alpha}_j}{4\pi d_j} - \frac{\vec{\alpha}_j}{4\pi \rho_j} \sum_{n=0}^{p} \mu_j^n P_n(cos \gamma)\right)\right|\right|
\end{equation}
Since $\mu_j = \frac{r}{\rho_j} \le \frac{a}{\rho-a}=\frac{1}{c}$, using the triangle inequality, the local expansion error will be bounded by:
\begin{equation}
E_{\vec{\psi}} \le \frac{1}{4\pi} \frac{\Gamma_{D_Q}}{(ca-a)}\left(\frac{1}{c}\right)^{p+1}
\end{equation}
where $\Gamma_{D_Q} = \sum_j \parallel \vec{\alpha_j}\parallel$.\\

\noindent Now suppose that $s$ vortices with strength $(\vec{\alpha}_j, j=1...s)$ are located inside a cubic box $J_l$ at level $l$ with center at $Q=(\rho,\alpha,\beta)$. The radius of the smallest sphere which encloses box $J_l$ is $a_{J_l} = \frac{\sqrt{3}}{2}W_{J_l}$ where $ W_{J_l}$ is the box width. Following the adaptive solver Scheme described in section \ref{}, an upper bound for $c$ is given by
\begin{equation}
c \geq \frac{2\sqrt{3}}{3}(n_D+1)-1,
\end{equation}

and for any $p \geq 1$, the local expansion error that results upon approximating the potential vector at any point $P$ within a box $I_l$ centered at the origin is bounded by:

\begin{equation}
E_{L.\vec{\psi}} \le \frac{1}{4\pi} \frac{\Gamma_{J_l}}{W_0} \frac{2_l}{n_D+1-\frac{\sqrt{3}}{2}}\left(\frac{1}{\frac{2\sqrt{3}}{3}(n_D+1)-1}\right)^{p+1}
\end{equation}

\subsection{Error bounds for the velocity vector induced by a singular source element}
Once again, let us consider a singular source element $j$ of strength $\vec{\alpha}_j$ located at $Q=(\rho_j,\theta_j,\varphi_j)$.
The velocity at a target element $P=(r,\theta,\varphi)$ is calculated as the curl of the potential vector $\vec{\psi}_j$
\begin{equation}
\vec{u}_j(P)=\nabla\times\vec{\psi}_j(P)=\nabla \times \left(\frac{\vec{\alpha}_j}{4\pi r}\sum_{n=0}^{\infty} \mu_j^n P_n(cos \gamma)\right)=\sum_{n=0}^{\infty} \frac{\rho_j^n}{4\pi} \nabla \left(\frac{P_n(cos \gamma)}{r^{n+1}}\right)\times\vec{\alpha}_j
\end{equation}
The error resulting from truncating $\vec{u}_j(P)$ at some order $p$ is:
\begin{equation}
E_{\vec{u}_j}=\left|\left|\sum_{n=p+1}^{\infty} \frac{\rho_j^n}{4\pi} \nabla \left(\frac{P_n(cos \gamma)}{r^{n+1}}\right)\times\vec{\alpha}_j\right|\right|
\end{equation}
Since $\parallel \vec{u}\times\vec{v}\parallel \le \parallel \vec{u} \parallel \parallel \vec{v} \parallel$ for any two vectors $\vec{u}$ and $\vec{v}$, using the triangle inequality, the truncating error will be bounded by:
\begin{equation}
E_{\vec{u}_j}\le \frac{\parallel\vec{\alpha}_j\parallel}{4\pi} \sum_{n=p+1}^{\infty} \rho_j^n \left|\left|\nabla \left(\frac{P_n(cos \gamma)}{r^{n+1}}\right)\right|\right|
\label{eq:error1}
\end{equation}
Since the norm of a vector does not vary when we change the reference frame, we will evaluate the expression of $\left|\left|\nabla \left(\frac{P_n(cos \gamma)}{r^{n+1}}\right)\right|\right|$ in a new frame obtained by rotating the old frame such that the source element $Q$ will be locatated on the new $zz'$ axe, that is we have $cos\gamma=cos\theta$
\begin{equation}
\nabla \left(\frac{P_n(cos \theta)}{r^{n+1}}\right)= -\frac{1}{r^{n+2}}\left(\left( n+1 \right)P_n(cos\theta), sin(\theta){P'}_n (cos\theta), 0 \right)^T
\end{equation}
\begin{equation}
\left|\left|\nabla \left(\frac{P_n(cos \theta)}{r^{n+1}}\right)\right|\right|=\frac{1}{r^{n+2}}\sqrt{(n+1)^2 P_n^2(cos\theta)+sin^2(\theta){P'}_n^2(cos\theta)}
\end{equation}
Knowing that 
\begin{equation}
(n+1)^2 P_n^2(x)+(1-x^2){P'}_n^2(x) \le (n+1)^2
\end{equation}
we obtain :
\begin{equation}
\left|\left|\nabla \left(\frac{P_n(cos \theta)}{r^{n+1}}\right)\right|\right| \le \frac{n+1}{r^n+2}
\end{equation}
Replacing in equation (\ref{eq:error1}), the multipole expansion error of the velocity vector is bounded by:
\begin{equation}
E_{\vec{u}_j}\le \frac{\parallel\vec{\alpha}_j\parallel}{4\pi r^2} \sum_{n=p+1}^{\infty} (n+1)\mu_j^n
\end{equation}
\begin{equation}
E_{\vec{u}_j}\le \frac{\parallel\vec{\alpha}_j\parallel}{4\pi r^2} \left[ \frac{\mu_j^{p+1}}{1-\mu_j}+\mu \frac{d}{d\mu_j} \left(\frac{\mu_j^{p+1}}{1-\mu_j}\right) \right]
\end{equation}
\begin{equation}
E_{\vec{u}_j}\le \frac{\parallel\vec{\alpha}_j\parallel}{4\pi r^2} \frac{\mu_j^{p+1}}{(1-\mu_j)^2}\left[p+2-\mu_j \left(p+1 \right) \right]
\end{equation}
\begin{equation}
E_{\vec{u}_j}\le \frac{1}{4\pi}\frac{\parallel\vec{\alpha}_i\parallel}{(r-\rho_j)^2}\left( \frac{\rho_j}{r}\right)^{p+1}\left[p+2-\frac{\rho_j}{r} \left(p+1 \right) \right]
\end{equation}
\\
Similary, when $r<\rho_j$, $\mu_j =\frac{r}{\rho_j}$, the velocity vector induced by $j$ at a target $P=(r,\theta,\phi)$ is calculated as the curl of the potential vector and has the following expression:
\begin{equation}
\vec{u}_j(P)=\nabla\times\vec{\psi}_j(P)=\nabla \times \left(\frac{\vec{\alpha}_j}{4\pi \rho_j}\sum_{n=0}^{\infty} \mu_j^n P_n(cos \gamma)\right)=\sum_{n=0}^{\infty} \frac{1}{4\pi \rho_j^n} \nabla \left( r^n P_n(cos \gamma)\right)\times\vec{\alpha}_j
\end{equation}
The error resulting from truncating $\vec{u}_j(P)$ at some order $p$ is:
\begin{equation}
E_{\vec{u}_j}=\left|\left|\sum_{n=p+1}^{\infty} \frac{1}{4\pi \rho_j^n} \nabla \left( r^n P_n(cos \gamma)\right)\times\vec{\alpha}_j\right|\right|
\end{equation}
\begin{equation}
E_{\vec{u}_j}\le \frac{\parallel\vec{\alpha}_j\parallel}{4\pi} \sum_{n=p+1}^{\infty} \frac{1}{\rho_j^n} \left|\left| \nabla \left( r^n P_n(cos \gamma)\right)\right|\right|
\label{eq:error2}
\end{equation}	
Once again, the norm of a vector does not vary when we change the reference frame. Calculating the norm in the new reference frame decribed above, we obtain:
\begin{equation}
\nabla \left( r^n P_n(cos \gamma)\right)= r^{n-1}(nP_n(cos\theta), -sin\theta P'_n (cos\theta),0)^T
\end{equation}
\begin{equation}
\left|\left| r^n P_n(cos \gamma)\right|\right| =r^{n-1}\sqrt{n^2 P_n^2(cos\theta)+sin^2\theta {P'}_n^2 (cos\theta)}
\end{equation}
Knowing that 
\begin{equation}
n^2 P_n^2(x)+(1-x^2){P'}_n^2(x) \le n^2
\end{equation}
we obtain :
\begin{equation}
\left|\left| r^n P_n(cos \gamma)\right|\right| =nr^{n-1}
\end{equation}
Replacing in equation (\ref{eq:error2}), the local expansion error of the velocity vector is bounded by:
\begin{equation}
E_{\vec{u}_j}\le \frac{\parallel\vec{\alpha}_j\parallel}{4\pi \rho_j} \sum_{n=p+1}^{\infty} n\mu_j^n
\end{equation}	
\begin{equation}
E_{\vec{u}_j}\le \frac{\parallel\vec{\alpha}_j\parallel}{4\pi \rho_j} \mu \frac{d}{d\mu_j} \left(\frac{\mu_j^{p+1}}{1-\mu_j}\right)
\end{equation}
\begin{equation}
E_{\vec{u}_j}\le \frac{\parallel\vec{\alpha}_j\parallel}{4\pi \rho_j^2} \frac{\mu_j^{p+1}}{(1-\mu_j)^2}\left[p+2-\mu_j p \right]
\end{equation}
\begin{equation}
E_{\vec{u}_j}\le \frac{1}{4\pi}\frac{\parallel\vec{\alpha}_i\parallel}{(\rho_j-r)^2}\left( \frac{r}{\rho_j}\right)^{p+1}\left[p+2-\frac{r}{\rho_j} p \right]
\end{equation}

\subsection{Error bounds for the velocity vector induced by a cluster of source elements}

Suppose that $s$ vortices with strength $(\vec{\alpha}_j, j=1...s)$ are located at the points $\vec{Q}_j = (\rho_j,\theta_j,\varphi_j)$ inside the sphere $D_Q$ of radius $a$ with center at $Q=(0,0,0)$, then at any $\vec{P}=(r,\theta,\varphi)$ with $r>a$, the velocity vector field is given by:

\begin{equation}
\vec{u}= \sum_j \vec{u}_j = \sum_j\sum_{n=0}^{\infty} \frac{\rho_j^n}{4\pi} \nabla \left(\frac{P_n(cos \gamma)}{r^{n+1}}\right)\times\vec{\alpha}_j
\end{equation}
Using the triangle inequality, and since $\mu_j \le \frac{a}{r}$, the multipole expansion error will be bounded by:
\begin{equation}
E_{\vec{u}} \le \sum_i E_{\vec{u}_j} \le \frac{1}{4\pi}\frac{\Gamma_{D_Q}}{(r-a)^2}\left( \frac{a}{r}\right)^{p+1}\left[p+2-\frac{a}{r} \left(p+1 \right) \right]
\end{equation}
where $\Gamma_{D_Q} = \sum_j \parallel \vec{\alpha_j}\parallel$.\\

\noindent Now suppose that $s$ vortices with strength $(\vec{\alpha}_j, j=1...s)$ are located inside a cubic box $J_l$ at level $l$ with center at $Q=(0,0,0)$. For any $p \geq 1$, the multipole expansion error that results upon approximating the velocity vector at any point $P$ outside the neighbours of $J_l$ is bounded by:
\begin{equation}
E_{M,\vec{u}}  \le  \frac{1}{4\pi}\frac{\Gamma_{J_l}}{W_0^2}\frac{4^{l^{*}}}{\left(n_D-\frac{\sqrt{3}-1}{2}\right)^2}\left(\frac{\sqrt{3}}{2 n_D+1}\right)^{p+1} \left[p+2-\frac{\sqrt{3}}{2n_D+1}(p+1)\right]
\end{equation}

Similary, suppose that $s$ vortices having  circulation $ (\vec{\alpha}_i, i=1...s)$ are located inside the sphere $D_Q$ of radius $a$ with center at $Q=(\rho,\alpha,\beta)$, and that $\rho = (c+1)a $ with $c >1 $, then for any target $P=(r,\theta,\phi)$  inside the sphere $D_0$ of radius $a$ centred at the Origin, the velocity vector is given by
\begin{equation}
\vec{u}= \sum_j \vec{u}_j = \sum_j \sum_{n=0}^{\infty} \frac{1}{4\pi \rho_j^n} \nabla \left( r^n P_n(cos \gamma)\right)\times\vec{\alpha}_j
\end{equation}
Since $\mu_j = \frac{r}{\rho_j} \le \frac{a}{\rho-a}=\frac{1}{c}$, using the triangle inequality, the local expansion error will be bounded by:
\begin{equation}
E_{\vec{u}} \le \sum E_{\vec{u}_j} \le \frac{1}{4\pi}\frac{\Gamma_{D_Q}}{(ca-c)^2}\left( \frac{1}{c}\right)^{p+1}\left[p+2-\frac{1}{c} p \right]
\end{equation}
where $\Gamma_{D_Q} = \sum_j \parallel \vec{\alpha_j}\parallel$.\\

\noindent Now suppose that $s$ vortices with strength $(\vec{\alpha}_j, j=1...s)$ are located inside a cubic box $J_l$ at level $l$ with center at $Q=(\rho,\alpha,\beta)$. For any $p \geq 1$, the local expansion error that results upon approximating the velocity vector at any point $P$ within a box $I_l$ centered at the origin is bounded by:

\begin{equation}
E_{L,\vec{u}} \le \frac{1}{4\pi} \frac{\Gamma_{J_l}}{W_0^2} \frac{ 4^{l^*}}{\left(n_D+1-\sqrt{3}\right)^2}\left( \frac{1}{\frac{2\sqrt{3}}{3}(n_D+1)-1}\right)^{p+1}\left[p+2-\left( \frac{1}{\frac{2\sqrt{3}}{3}(n_D+1)-1}\right) p \right]
\end{equation}

\subsection{Regularization error bounds for the potential vector}

In the case of regularized vortex method, an additional error $E_{\sigma,\vec{\psi}} $ aroses since $G_{\sigma}$ deviates from $\frac{1}{4\pi r}$. Let us consider $s$ vortices with strength $(\vec{\alpha}_j, j=1...s)$  located inside a cubic box $J_l$. Introducing the kernel $\beta_{\sigma}(r) = \frac{1}{\sigma} \beta \left( \frac{r}{\sigma}\right)$, and  $\beta (r) =\abs{G(r)-\frac{1}{4\pi r} } $ we obtain:

\begin{align}
E_{\sigma,\vec{\psi}} & =  \norm{ \sum_{j \in J_l} \vec{\alpha}_j \left(G_{\sigma}(r_j)-\frac{1}{4\pi r_j} \right)} \nonumber \\
                   & \le \sum_{j \in J_l} \lVert \vec{\alpha}_j \rVert  \abs{G_{\sigma}(r_j)-\frac{1}{4\pi r_j} } \nonumber \\
                   & \le \abs{G_{\sigma}(r^*)-\frac{1}{4\pi r^*} } \sum_{j \in J_l} \lVert \vec{\alpha}_j \rVert   \nonumber \\
                   & \le \Gamma_{J_l} \beta_{\sigma}(r^*)
\end{align}

\noindent where $\Gamma_{J_l} = \sum_{j} \lVert \vec{\alpha}_j \lVert$, $r_j$ is the distance between source element $j$ and $P$, $r^*$ is the distance that maximizes the kernel $\beta_{\sigma}(r)$. $r^*$ depends only on the choice of the core function and for most cases we have $r^* =(n_D+0.5)W_{min}=(n_D+0.5)\frac{W_0}{2^l}$.

\subsection{Regularization error bounds for the velocity vector}
The regularizing error for the velocity arose since $q_\sigma(r)$ deviates from $\frac{1}{4\pi}$. Let us consider $s$ vortices with strength $(\vec{\alpha}_j, j=1...s)$  located inside a cubic box $J_l$. For any target element $i$ we have:

\begin{align}
E_{\sigma,\vec{u}} & = \norm{ \sum_{j \in J_l} \frac{1}{4\pi r_j^3}(\vec{x_i}-\vec{x}_j)\wedge\vec{\alpha_j} - \frac{q_{\sigma}(r_j)}{ r_j^3}(\vec{x_i}-\vec{x}_j)\wedge\vec{\alpha_j} } \nonumber \\
& \leq \sum_{j \in J_l} \norm{\frac{1}{4\pi r_j^3}(\vec{x_i}-\vec{x}_j)\wedge\vec{\alpha_j} - \frac{q_{\sigma}(r_j)}{ r_j^3}(\vec{x_i}-\vec{x}_j)\wedge\vec{\alpha_j}} \nonumber \\
& \leq \sum_{j \in J_l}  \frac{1}{4\pi r_j^3}\arrowvert 1-4\pi q_{\sigma}(r_j) \arrowvert \parallel(\vec{x}_i-\vec{x}_j)\wedge\vec{\alpha_j} \parallel \nonumber \\
& \leq \sum_{j \in J_l}  \frac{1}{4\pi r_j^2}\arrowvert 1-4\pi q_{\sigma}(r_j) \arrowvert \parallel\vec{\alpha_j} \parallel \nonumber \\
& \leq  \frac{\Gamma_{J_l}}{4\pi {r^*}^2}\arrowvert 1-4\pi q(\frac{r^*}{\sigma}) \arrowvert \nonumber \\
& \leq  \Gamma_{J_l} \kappa_{\sigma}(r^*)
\end{align}

\noindent where $\Gamma_{J_l} = \sum_{j} \lVert \vec{\alpha}_j \lVert$, $\kappa (r)= \frac{1}{4\pi r^2}\arrowvert 1-4\pi q(r) \arrowvert $ and $\kappa_{\sigma}(r)=\frac{1}{\sigma^2}\kappa (\frac{r}{\sigma})$, $r_j$ is the distance between source element and target elements, $r^*$ is the distance that maximizes the kernel $\kappa_{\sigma}(r)$. 

\end{document}